\documentclass[onecolumn,10pt,journal,compsoc]{IEEEtran}
\ifCLASSOPTIONcompsoc
  \usepackage[nocompress]{cite}
\else
  \usepackage{cite}
\fi
\ifCLASSINFOpdf
\else
\fi
\usepackage{amsmath,graphicx}
\usepackage{amsfonts}
\usepackage{epsfig,amssymb,latexsym,graphics,float}
\usepackage{setspace,subfig}
\usepackage{citesort}
\usepackage{amsopn}
\usepackage{amssymb,times}
\usepackage{float}
\usepackage{url}
\usepackage[ruled,vlined,linesnumbered]{algorithm2e}
\begin{document}
\title{State Space Model based Trust Evaluation \\over Wireless Sensor Networks: \\
An Iterative Particle Filter Approach}
\author{Bin~Liu,
        and~Shi~Cheng 
\IEEEcompsocitemizethanks{\IEEEcompsocthanksitem B. Liu is with the School of
Computer Science and Technology, Nanjing University of Posts and Telecommunications, Nanjing,
210023, China.\protect\\
E-mail: bins@ieee.org
\IEEEcompsocthanksitem S. Cheng is with School of Computer Science, Shaanxi Normal University, Xi’an, 710062 China}
\thanks{Manuscript received Dec. 16, 2016; accepted March 9, 2017.}}
\IEEEtitleabstractindextext{%
\begin{abstract}
In this paper we propose a state space modeling approach for trust evaluation in wireless sensor networks.
In our state space trust model (SSTM), each sensor node is associated with a trust metric, which measures to what extent
the data transmitted from this node would better be trusted by the server node.
Given the SSTM, we translate the trust evaluation problem to be a nonlinear state filtering problem.
To estimate the state based on the SSTM, a component-wise iterative state inference procedure is proposed to work in tandem with the particle filter,
and thus the resulting algorithm is termed as iterative particle filter (IPF).
The computational complexity of the IPF algorithm is theoretically linearly related with the dimension of the state. This property is desirable
especially for high dimensional trust evaluation and state filtering problems.
The performance of the proposed algorithm is evaluated by both simulations and real data analysis.
\end{abstract}
\begin{IEEEkeywords}
trust model, wireless sensor network, trust evaluation, iterative particle filter, fault detection.
\end{IEEEkeywords}}
\maketitle
\IEEEdisplaynontitleabstractindextext
\IEEEpeerreviewmaketitle
\IEEEraisesectionheading{\section{Introduction}\label{sec:intro}}
\IEEEPARstart{W}{ireless} sensor networks (WSN) are networked systems that consist of autonomous nodes collaborating to perform an application task.
The nodes of a networked system are usually spatially distributed and equipped with limited sensing, computing and communication capabilities.
The research on WSN has gained significant concern in the last decade. The related application domains include but are not limited to
health care \cite{otto2006system}, energy security \cite{leon2007application}, environmental monitoring \cite{werner2006deploying,liu2015toward} and
military information integration \cite{lee2009wireless,diamond2007application}.

The performance of WSN depends on collaboration among distributed sensor nodes, while those nodes are often unattended with severe energy constraints
and limited reliability. In such conditions, it is important to evaluate the trustworthiness of participating nodes since
trust is the major driving force for collaboration. The focus of this paper is to propose a state space trust model (SSTM)
along with a corresponding trust evaluation algorithm in the context of WSN.

The research on trust evaluation has been extensively performed in the context of several diverse domains such as security \cite{blaze1999role,blaze1996decentralized},
electronics commerce \cite{josang2007survey,resnick2002trust}, peer-to-peer networks \cite{selcuk2004reputation,singh2003trustme},
and ad hoc and sensor networks \cite{michiardi2002core,ganeriwal2008reputation,liu2015toward,wang2017online}. The main objective of the trust evaluation module is to expose an
output metric that can be used as a representative of the subjective expectation of the sensor nodes' future behaviors.
This trust metric can be used in several ways. For example, the trust value of each node can be used as a weight for a data reading reported by this node.
Then the data fusion can be performed on these weighted data readings, thereby reducing the impact of untrustworthy nodes \cite{ganeriwal2008reputation}.
In addition, the evolution of trust over time can facilitate on-line detection of misbehaving nodes. Last but not least, the trust value
can be used as a decision making criteria for the end-user to take appropriate measures such as replacing detected faulty nodes.
Although various trust models and trust evaluation approaches are available
\cite{gambetta2000can,theodorakopoulos2004trust,josang1999algebra,manchala1998trust,resnick2002trust,jsang2002beta,josang2007dirichlet,nielsen2007bayesian},
there are still many challenges that need
to be addressed. It is not clear what are the fundamental rules the trust models must follow, therefore there is neither a consensus on the definition of trust,
nor a common rule for specifying an appropriate trust metric for a given problem. As a result, the design of
trust models is still at the empirical stage.

Motivated by the lack of a unified theoretical framework to build up trust models, in this paper, we introduce a generic theoretical model, namely SSTM, in the context of WSN data analysis. We also propose a novel trust evaluation algorithm, termed iterative particle filter (IPF), based on the framework of SSTM. We show that the SSTM framework is extensible and generic, and can include related existing trust evaluation approaches, e.g., the Bayesian dynamic model based particle filter (BDMPF) \cite{liu2015toward}, as a special case.

The remainder of the paper is organized as follows. In section 2, we describe the SSTM. In section 3,
we introduce the proposed IPF algorithm in detail. In section 4, we report the simulation and real data analysis results in applying IPF for
trust evaluations over WSN. Finally, in section 5, we conclude this paper.
\section{Network topology model}
\begin{figure}[h]
\begin{tabular}{c}
\centerline{\includegraphics[width=4in,height=2in]{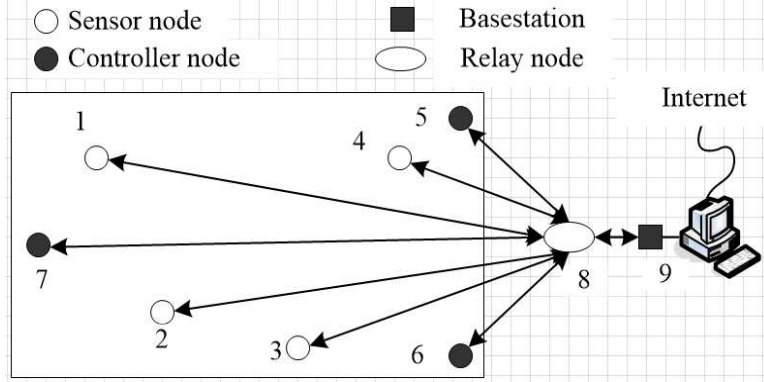}}
\end{tabular}
\caption{The WSN network topology model under consideration}\label{network_topology}
\end{figure}

We focus on the network topology model as shown in Fig.\ref{network_topology}. This model was considered in \cite{liu2015toward}.
The sensor nodes are arranged to sense the environmental parameters and report them to the relay node in real time.
The relay node receives the sensor readings from the sensor nodes, and then sends them to a basestation that is
communicated with a server computer node. All the sensor readings are gathered and analyzed at the server computer node.
The server computer node is connected with Internet, such
that the result of real-time data analysis can be checked remotely by the end-user of the WSN system.
Every sensor reading consists of the sensed environmental parameter values and the corresponding sensor ID.
Therefore, at the server computer node, we can easily find out the corresponding source senor node for each sensor reading.
The controller nodes receive feedback signal from the computer node, and then control several apparatus in order to tune
the environmental parameters.
\section{State space trust model (SSTM)}\label{sec:SSTM}
In this Section, we describe the SSTM in detail. We show that, based on the SSTM,
trust evaluation over WSN can be formulated as a nonlinear state filtering problem.

In SSTM, the trustworthiness of each sensor node is modeled by a trust index,
to measure to what extent the data transmitted from
this node would better be trusted by the server computer node.
The state vector $x_k$ in the SSTM is defined as follows
\begin{equation}
x_k\triangleq[x_{k,1},x_{k,2},\ldots,x_{k,d}]
\end{equation}
where the element $x_{k,j}$ denotes the trust value of the $j$th node at the $k$th time step, and $d$ is the dimension of $x_k$,
corresponding to the number of sensor nodes under
consideration. The value space of the trust metric $x_{i,j}$ is $[0,1]$, whereby the extreme value 1 means ``fully trusted", and 0
indicates the opposite, that is ``totally un-trusted".

The trust propagation law over time is modeled by the aging mechanism as follows \cite{ganeriwal2008reputation,buchegger2003coping}
\begin{equation}\label{eqn:state_trans}
x_{k+1}=\alpha x_k+v,\quad v\sim\mathcal{N}(0,Q),
\end{equation}
where $0<\alpha<1$ denotes the aging parameter, $Q$ denotes a diagonal matrix and $\mathcal{N}(0,Q)$ is a zero-mean
Gaussian distribution with covariance $Q$.
In \cite{ganeriwal2008reputation}, the value of $\alpha$ is set to be 0.95.
The value of $\alpha$ can also be found out by comparing the evolution of
the trust in a system with and without aging weight, respectively \cite{buchegger2003coping}. Note that if the value of $x_{k+1}$ obtained with Eqn.\ref{eqn:state_trans} falls outside of the range [0,1], then we draw a new value of $x_{k+1}$ via Eqn.(\ref{eqn:state_trans}) until it falls within the range [0,1].

A \emph{generative} model of the sensor readings, which relates the trust metric with the real sensor readings, is specified by the
likelihood function. Let $y_k$ denote the data collected by the server computer node at the $k$th time step.
We have $y_k=[y_{k,1},y_{k,2},\ldots,y_{k,d}]$, where $y_{k,j}$
denotes the data reported by the $j$th node at the $k$th time step.
The likelihood function is designed to be
\begin{equation}\label{likelihood}
p(y_k|x_k)=\exp\left(\frac{-\sum_{j=1}^d|x_{k,j}-V\left(\{x_{k,n}\}_{n\in\{1:d\}\backslash j},y_k,j\right)|}{\beta}\right),
\end{equation}
where $\{1:d\}\backslash j$ denotes $\{1,\ldots,j-1,j+1,\ldots,d\}$, $|A|$ denotes the absolute value of $A$, $0<\beta<1$ is a
free parameter and
\begin{equation}\label{eqn:voting}
V\left(\{x_{k,n}\}_{n\in\{1:d\}\backslash j},y_k,j\right)\triangleq \frac{\sum_{n\in\{1:d\}\backslash j}x_{k,n}U(n,j,y_k)}{\sum_{n\in\{1:d\}\backslash j}x_{k,n}}.
\end{equation}
Eqn.(\ref{eqn:voting}) describes the computation of the voting metric of the $j$th node, given by the other nodes.
The item $U(n,j,y_k)$ in Eqn.(\ref{eqn:voting}) denotes the voting result node $n$ gives to $j$, and is defined to be
\begin{equation}
U(n,j,y_k)=\left\{\begin{array}{ll}
1,\quad\mbox{if}\quad|y_{k,n}-y_{k,j}|< r \\
0,\quad\mbox{otherwise} \end{array} \right.
\end{equation}
where $r$ denotes a preset threshold for determining whether a pair of nodes reports sensor readings with permissible
differences.
The underlying assumption adopted in defining $U(n,j,y_k)$ is that,
sensor readings reported by mutually trusted sensor nodes should not have significant difference.
This assumption is reasonable for many WSN applications, wherein the nodes are mutual spatial neighbors among with each other
and the trusted sensor readings should be spatially correlated with each other.
This assumption is also
in analogy with the social trust, wherein mutually trusted social entities report similar opinions (data)
on an object or event in an ad hoc context.
In Eqn.(\ref{eqn:voting}), $U(n,j,y_k)$ is weighted
by $\frac{x_{k,n}}{\sum_{i\in\{1:d\}\backslash j}x_{k,i}}$, for each $n\in\{1:d\}\backslash j$.
In such a way, the impact of each node is adjusted according to its trustworthiness, and thus the impact of
untrustworthy nodes is reduced, in generating the final voting metric of node $j$, i.e., $V\left(\{x_{k,n}\}_{n\in\{1:d\}\backslash j},y_k,j\right)$

Given the data collected by the server computer node until the $k$th time step, denoted
by $y_{1:k}\triangleq\{y_1, \ldots, y_k\}$,
we are right now concerned with
the calculation of the \emph{posterior} probability density function (pdf) $\pi_k=p(x_k|y_{1:k})$
in a Bayesian inference framework.

According to Bayesian philosophy \cite{berger2013statistical}, given $y_{1:k}$, all the information on $x_k$ is encoded by
the \emph{posterior}, as long as the prior pdf and the likelihood function are specified appropriately.
Particle filters (PFs), a.k.a. Sequential Monte Carlo methods, are recognized as a general approach to address such a Bayesian state estimation
problem \cite{gordon1993novel,arulampalam2002tutorial,liu2010multi}. In comparison with other state filtering algorithms, such as the Kalman filter and its variants,
PFs have striking advantages in coping with nonlinearities and/or non-Gaussian noises in the model \cite{arulampalam2002tutorial,smith2013sequential,liu2008particle}.
However, as a Monte Carlo method, the PF algorithm inevitably suffers from the well-known curse of dimensionality,
that is the PF may collapse in case of high dimensional state vector \cite{snyder2008obstacles,bengtsson2008curse,rebeschini2015can}.

Regarding our problem at hand, the dimension of the state, i.e., $d$, is equal to the number of sensor nodes under consideration. If the network of our concern
consists of massive sensor nodes densely arranged, the corresponding state will become high dimensional, thus the conventional PF algorithms may become
invalid. We propose in Section \ref{sec:IPF} a novel PF algorithm, namely IPF, to get around of the above computation problem caused by high dimensionality.
\section{Iterative Particle Filter}\label{sec:IPF}
In this Section, we introduce the proposed IPF algorithm in detail.
To begin with, we give a brief review on a conventional PF algorithm, termed bootstrap PF, to fix the notations.
\subsection{Bootstrap particle filter}\label{sec:bootstrap}
The bootstrap PF is a general practical nonlinear state filter, which typically proceeds by Monte Carlo approximation.
This algorithm has a recursive structure in its implementation, thus it allows the state filter to be computed on-line over a long time horizon.
The recursion is at the level of probability measures, and the target distribution $\pi_k$ is approximated by the empirical
distribution $\hat{\pi}_k$. The distribution $\hat{\pi}_k$ is then computed by the recursion
\begin{equation}
\hat{\pi}_0=\pi_0, \quad \hat{\pi}_k=\mbox{F}_k\hat{\pi}_{k-1}
\end{equation}
where $\pi_0$ denote a \emph{prior} belief on the state, and $\mbox{F}_k$ denotes an operator that consist of two steps:
\begin{equation}
\hat{\pi}_{k-1}\xrightarrow{\mbox{Prediction}}\hat{\pi}_{k-}\xrightarrow{\mbox{Correction}}\hat{\pi}_{k}.
\end{equation}
The empirical distribution $\hat{\pi}_{k-1}$ is the output of the algorithm at the $k-1$th $(k>1)$ time step, and is represented as
\begin{equation}
\hat{\pi}_{k-1}=\frac{1}{N}\sum_{i=1}^N\sigma_{x_{k-1}^i},
\end{equation}
where $N\geq 1$ is the number of particles used in the algorithm,
$(x_{k-1}^i)_{i=1,\ldots,N}$ are independent identically distributed (i.i.d.) samples from $\hat{\pi}_{k-1}$,
and $\sigma_x$ denotes the delta function located at $x$.

In the prediction step, a set of new particles $\{x_{k-}^i\}_{i=1,\ldots,N}$ is generated according to the state transition law,
i.e., Eqn.(\ref{eqn:state_trans}) for the problem of our concern. Specifically, we have
\begin{equation}
x_{k-}^i=\alpha x_{k-1}^i+v, v\sim\mathcal{N}(0,Q), i=1,\ldots,N.
\end{equation}
Note that the value of $x_{k-}^i$ needs to be bounded within $[0,1]$.
Provided that its value jumps outside of the bounded space $[0,1]$,
we just generate a new value for $x_{k-}^i$ using Eqn.(\ref{eqn:state_trans}).

These particles $\{x_{k-}^i\}_{i=1,\ldots,N}$ are then weighted in the correction step.
The weights are termed importance weights in the context of PF, and
are calculated as follows
\begin{equation}\label{eqn:weights}
w^i=\frac{\bar{w}^i}{\sum_{i=1}^N\bar{w}^i}, i=1,\ldots,N,
\end{equation}
where
\begin{equation}
\bar{w}^i=p(y_k|x_{k-}^i), i=1,\ldots,N.
\end{equation}
Then let $\hat{\pi}_k=\sum_{i=1}^Nw^i\delta_{x_{k-}^i}$.
In bootstrap PF, a resampling procedure is included to prevent the phenomenon of particle degeneracy,
that is more and more particles get zero weights and are lost.
The basic operations of resampling are described in Algorithm \ref{algo:resampling}.
\begin{algorithm}[htbp]
For $i=1, \ldots, N$, sample an index $j(i)$ distributed according to the discrete distribution
with $N$ elements satisfying Pr$\{j(i)=l\}=w^i$\;
For $i=1, \ldots, N$, let $x_{k}^i=x_{k-}^{j(i)}$, $w_i=1/N$\;
Output $\{x_{k}^i\}_{i=1,\ldots,N}$.
  \caption{\label{algo:resampling}Resampling algorithm with input $\{x_{k-}^i,w^i\}_{i=1,\ldots,N}$}
\end{algorithm}

A summarization of the bootstrap PF is described in Algorithm \ref{algo:bootstrap},
where $K$ denotes the total number of time steps under consideration.

\begin{algorithm}[H]
  Let $\hat{\pi}_0=\pi_0$\;
  \For{$k=1,\ldots, K$ }{
    Sample i.i.d. $x_{k-1}^i$, $i=1,\ldots,N$ from the distribution $\hat{\pi}_{k-1}$\;
    Sample $x_{k-}^i\sim p(x_k|x_{k-1}^i), i=1,\ldots,N$ using Eqn.(\ref{eqn:state_trans})\;
    Calculate the importance weights $w^i, i=1,\ldots,N$, using Eqn.(\ref{eqn:weights})\;
    Run \textbf{Algorithm \ref{algo:resampling}} with input $\{x_{k-}^i,w^i\}_{i=1,\ldots,N}$,
    and get $\{x_k^i\}_{i=1,\ldots,N}$\;
    Let $\hat{\pi}_k=\frac{1}{N}\sum_{i=1}^N\sigma_{x_k^i}$\;
  }
  \caption{\label{algo:bootstrap}Bootstrap particle filter}
\end{algorithm}

It is shown that the empirical distribution $\hat{\pi}_k$ converges to the exact target
distribution $\pi_k$ as $N\rightarrow\infty$ \cite{hu2008basic,crisan2002survey}.
We refer to \cite{arulampalam2002tutorial} for a detailed overview of PF algorithms and
the related analysis.
\subsection{Derivation of the IPF algorithm}
Since conventional PF algorithms, such as the bootstrap PF presented in Sec.\ref{sec:bootstrap},
have inevitable drawbacks in dealing with filtering problems with high dimensional state
vectors \cite{snyder2008obstacles,bengtsson2008curse,rebeschini2015can}, here we derive a novel IPF algorithm,
based on the specific structure of our model, to get around of
the obstacles resulted from high dimensionality.

Observe that the likelihood function constructed in Eqn.(\ref{likelihood}) can be factorized as follows
\begin{equation}\label{factorized_liklihood}
p(y_k|x_k)=\prod_{j=1}^d\exp\left(\frac{-\mid x_{k,j}-V\left(\{x_{k,n}\}_{n\in\{1:d\}/j},y_k,j\right)\mid}{\beta}\right).
\end{equation}
Therefore, conditional on $\{x_{k,n}\}_{n\in\{1,\ldots,d\}/j}$,
we can calculate the likelihood of $x_{k,j}$ as follows
\begin{equation}\label{eqn:component_lik}
p(y_k|x_{k,j}|\{x_{k,n}\}_{n\in\{1,\ldots,d\}/j})=\exp\left(\frac{-\mid x_{k,j}-V\left(\{x_{k,n}\}_{n\in\{1:d\}/j},y_k,j\right)\mid}{\beta}\right).
\end{equation}
Based on Eqn.(\ref{eqn:state_trans}), the state transition law of $x_{k,j}$ can be shown to be
\begin{equation}\label{eqn:component_state_trans}
x_{k+1,j}=\alpha x_{k,j}+v_j, v_j\sim\mathcal{N}(0,Q_{jj}),
\end{equation}
where $Q_{j,j}$ denotes the $j$th diagonal element of matrix $Q$.

Given the component-wise likelihood and state transition function, specified by Eqns.(\ref{eqn:component_lik}) and
(\ref{eqn:component_state_trans}), respectively, we can accordingly calculate the component-wise \emph{posterior}, which is only a
one-dimensional distribution and thus is very easy to be sampled from.

The basic idea of IPF is that, at each time step, instead of sampling straightforwardly from the
high-dimensional \emph{posterior} (such as in conventional
PF), we perform component-wise inferences by sampling from a set of component-wise \emph{posterior} pdfs,
and then update the estimate of
the trust iteratively.
Specifically, the component-wise inference operations are described in Algorithm \ref{algo:iteration_IPF},
wherein $\|A-B\|$ denotes the Euclidean distance between the two vectors $A$ and $B$, and $T_x$ denotes a preset threshold for
determining if values of a pair of trust vectors have significant difference with each other.

Finally, the IPF is described in Algorithm \ref{algo:IPF}.
Although the proposed IPF algorithm has an iterative component, our experiments in Section \ref{sec:experiments} (see Fig.\ref{fig:convergence_IPF}) show that it needs just a few iterations in order to converge.
\begin{algorithm}[!htb]
Input: $\hat{\pi}_{k-1}$\;
Initialize $X_o$ to be a $d$ dimensional vector with all elements being 0\;
Let $m=1$ ($m$ denotes the iteration index)\;
     \While{$m=1$ or $\sqrt{\|\hat{x}_k-X_o\|/d}>T_x$ ($T_x$ denotes a preset threshold)}{
        \If{$m>1$}{
            Let $X_o=\hat{x}_k$\;
            }
        \For{$j=1,\ldots, d$ }{
            Sample i.i.d. $x_{k-1,j}^i$, $i=1,\ldots,N$ from the distribution $\hat{\pi}_{k-1,j}$\;
            Sample $x_{k-,j}^i\sim p(x_{k,j}|x_{k-1,j}^i), i=1,\ldots,N$ using Eqn.(\ref{eqn:component_state_trans})\;
            Calculate the importance weights $\bar{w}^i=p(y_k|x_{k-,j}^i|\{\hat{x}_{k,n}\}_{n\in\{1,\ldots,d\}/j}), i=1,\ldots,N$, using Eqn.(\ref{eqn:component_lik})\;
            Normalize the importance weights by $w^i=\bar{w}^i/\sum_{u=1}^N\bar{w}^u, i=1,\ldots,N$\;
            Run \textbf{Algorithm \ref{algo:resampling}} with input $\{x_{k-,j}^i,w^i\}_{i=1,\ldots,N}$,
            and get $\{x_{k,j}^i\}_{i=1,\ldots,N}$\;
            Let $\hat{\pi}_{k,j}=\frac{1}{N}\sum_{i=1}^N\sigma_{x_{k,j}^i}$\;
            Update the $j$th dimension of $\hat{x}_k$ by $\hat{x}_{k,j}=\frac{1}{N}\sum_{i=1}^Nx_{k,j}^i$\;
        }
        Let m=m+1\;
     }
     Output $\hat{\pi}_{k}$ and $\hat{x}_{k}$.
  \caption{\label{algo:iteration_IPF}Iterative component-wise inference within the IPF at time step $k$}
\end{algorithm}
\begin{algorithm}[!htb]
  Let $\hat{\pi}_0=\pi_0$\;
  \For{$k=1,\ldots, K$ }{
     Run Algorithm \ref{algo:iteration_IPF} with input $\hat{\pi}_{k-1}$. Output $\hat{\pi}_{k}$ and $\hat{x}_{k}$.
  }
  \caption{\label{algo:IPF}The proposed iterative particle filter algorithm}
\end{algorithm}
\subsection{Connections to existing work}\label{sec:connection}
The proposed IPF algorithm has a close connection to the BDMPF algorithm \cite{liu2015toward}.
Both algorithms are developed within the Bayesian state filtering framework, while their essential difference
lies in the design of the model.
In BDMPF, the voting metric of the $j$th node, given by the other nodes, is computed as follows
 \begin{equation}\label{eqn:voting_BDMPF}
V\left(\{x_{k,n}\}_{n\in\{1:d\}\backslash j},y_k,j\right)\triangleq \frac{\sum_{n\in\{1:d\}\backslash j}U(n,j,y_k)}{d-1}.
\end{equation}
In comparison with Eqn.(\ref{eqn:voting}), we see that Eqn.(\ref{eqn:voting_BDMPF}) is equivalent to Eqn.(\ref{eqn:voting})
in case of $x_{k,n}=1$ for any $n\in\{1:d\}\backslash j$. In another word, in calculating the voting metric of node $j$,
BDMPF assumes that all the other sensor nodes are all completely trusted. Clearly such an assumption is easy to be violated in practice.

In addition, regarding BDMPF and IPF, the difference in the their model structures leads to a corresponding difference in the
related inference algorithms. In the inference process, the IPF algorithm employs the fact that $x_{k,1}, x_{k,2}, \ldots, x_{k,d}$
are correlated with each other and thus should be
estimated jointly, while the BDMPF assumes that $x_{k,1}, x_{k,2}, \ldots, x_{k,d}$ are independent with each other,
thus are estimated separately.

Therefore, the proposed IPF algorithm is preferable to BDMPF for WSN applications, wherein the trustworthy sensor readings
are statistically correlated with each other and are independent with those yielded by un-trusted nodes.
The empirical results presented in Section \ref{sec:experiments} are consistent with the above analysis.
\section{Performance Evaluation}\label{sec:experiments}
In this section, we present performance evaluation results of the proposed IPF
algorithm based on simulations and real data analysis.
\subsection{Simulation results}
We tested our algorithm based on the simulation case that was used in
\cite{liu2015toward}.
\subsubsection{Simulation setting}\label{sec:simulation_setting}
In this case, we have ten sensor nodes involved, each of which reports its sensor reading to the server computer node at 100 discrete time steps.
The network topology related with this simulation case is the same as shown in Fig.\ref{network_topology}.
The values of trustworthy sensor readings are simulated to be normally distributed centering at 20 degrees Celsius at each time step.
Among the sensor nodes, seven of them are trustworthy as they transmit normal sensor readings from beginning to end.
The remaining nodes, indexed by ``Sensor A", ``Sensor B" and ``Sensor C", have different types of
unreliability in their behavior.
Specifically, ``Sensor A" is simulated to be unreliable from 31st to 70th time steps, during which the sensor reading value it transmits
rises gradually from 20 to 40
degrees Celsius between the 30th and 50th time step, and then falls back gradually to 20 degrees
Celsius between the 50th and 70th time step. The sensor reading of ``Sensor B" is simulated to be uniformly distributed
between 0 and 100 degrees Celsius at each time step. ``Sensor C" is simulated to work normally from 1st to 50th time step and then
stop reporting any values afterwards. This phenomenon is termed as ``Sleeper attack" in \cite{ganeriwal2008reputation}.
\subsubsection{Performance comparison with the BDMPF algorithm}
We compared our IPF algorithm with the BDMPF algorithm proposed in \cite{liu2015toward} by Monte Carlo simulations.
We ran 100 times of independent Monte Carlo runs of the IPF algorithm and the BDMPF algorithm.
These two algorithms were initialized by the same parameter setting as shown in Table 1.
At the beginning, the trust metric of every sensor node was set to be 0.5.
\begin{table}[H]
\caption{Parameter setting of the IPF algorithm in the Monte Carlo simulation test}
\begin {center}
\begin{tabular}{c|c|c|c|c|c}
\hline\hline
$N$ & $\alpha$ & $Q$ & $\beta$ & $r$ & $T_x$\\
\hline
$100$ & $0.85$ & diag$([0.01,\ldots,0.01])$ & $0.1$ & $0.6$ & $1e-5$ \\
\hline\hline
\end{tabular}
\end {center}
\end{table}

\begin{figure}[H]
\begin{tabular}{c}
\centerline{\includegraphics[width=3in,height=2.2in]{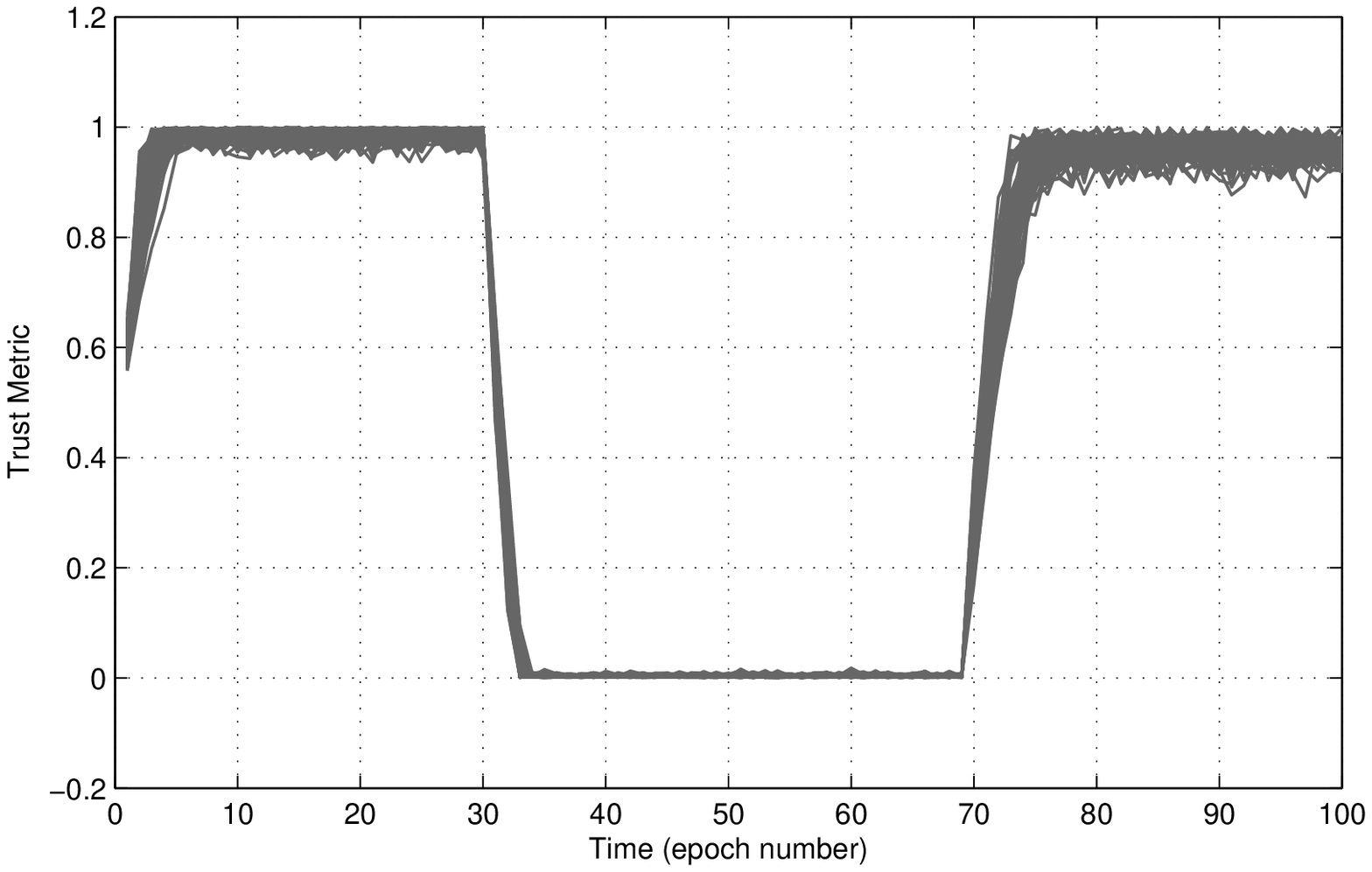}\includegraphics[width=3in,height=2.2in]{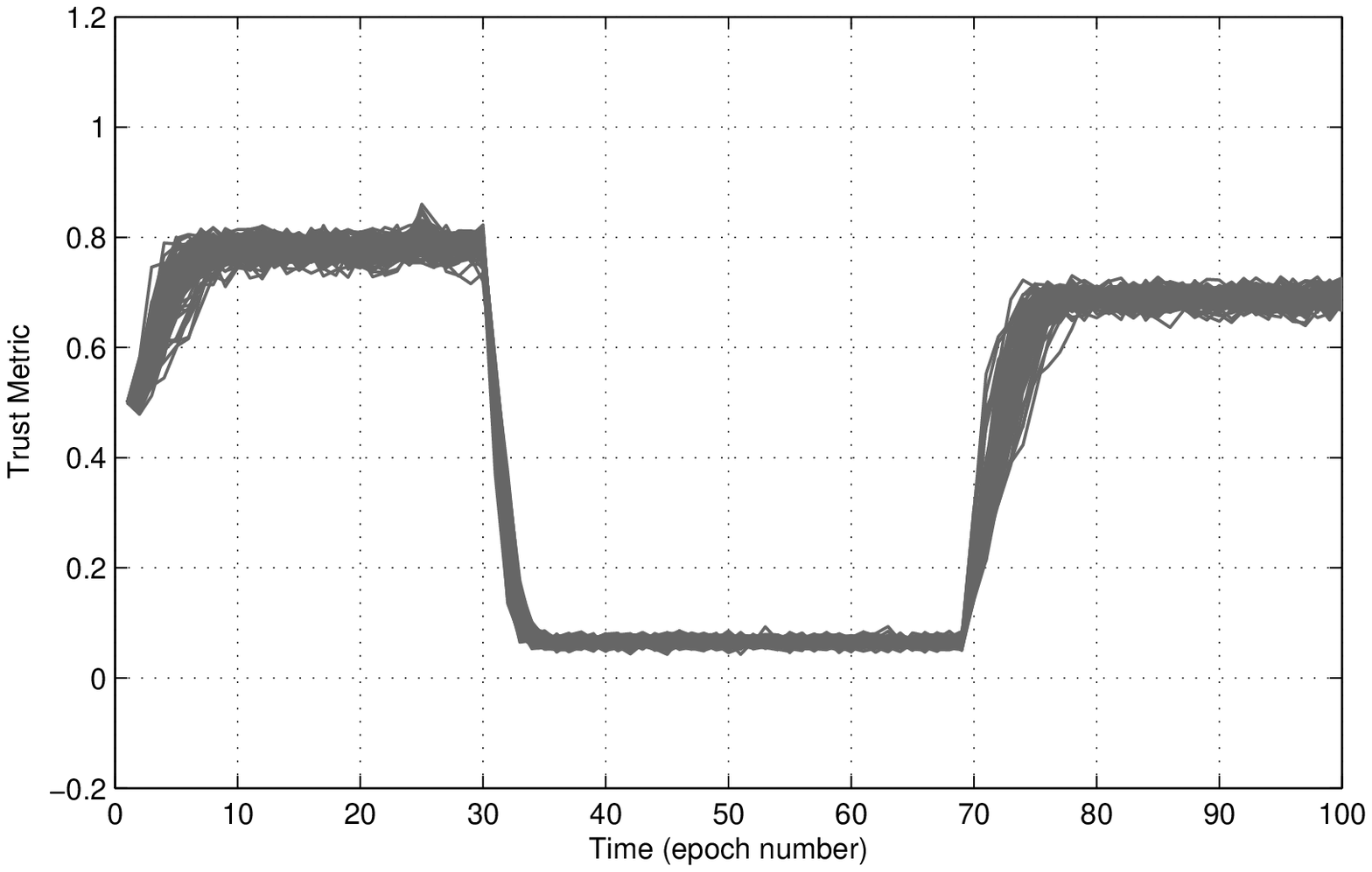}}
\end{tabular}
\caption{Left: Traces of the estimated trust metric of ``Sensor A" in 100 independent
Monte Carlo runs of the IPF algorithm. Right: Traces of the estimated trust metric of ``Sensor A" in 100 independent
Monte Carlo runs of the BDMPF algorithm.} \label{fig:Node_A_trust}
\end{figure}

\begin{figure}[H]
\begin{tabular}{c}
\centerline{\includegraphics[width=3in,height=2.2in]{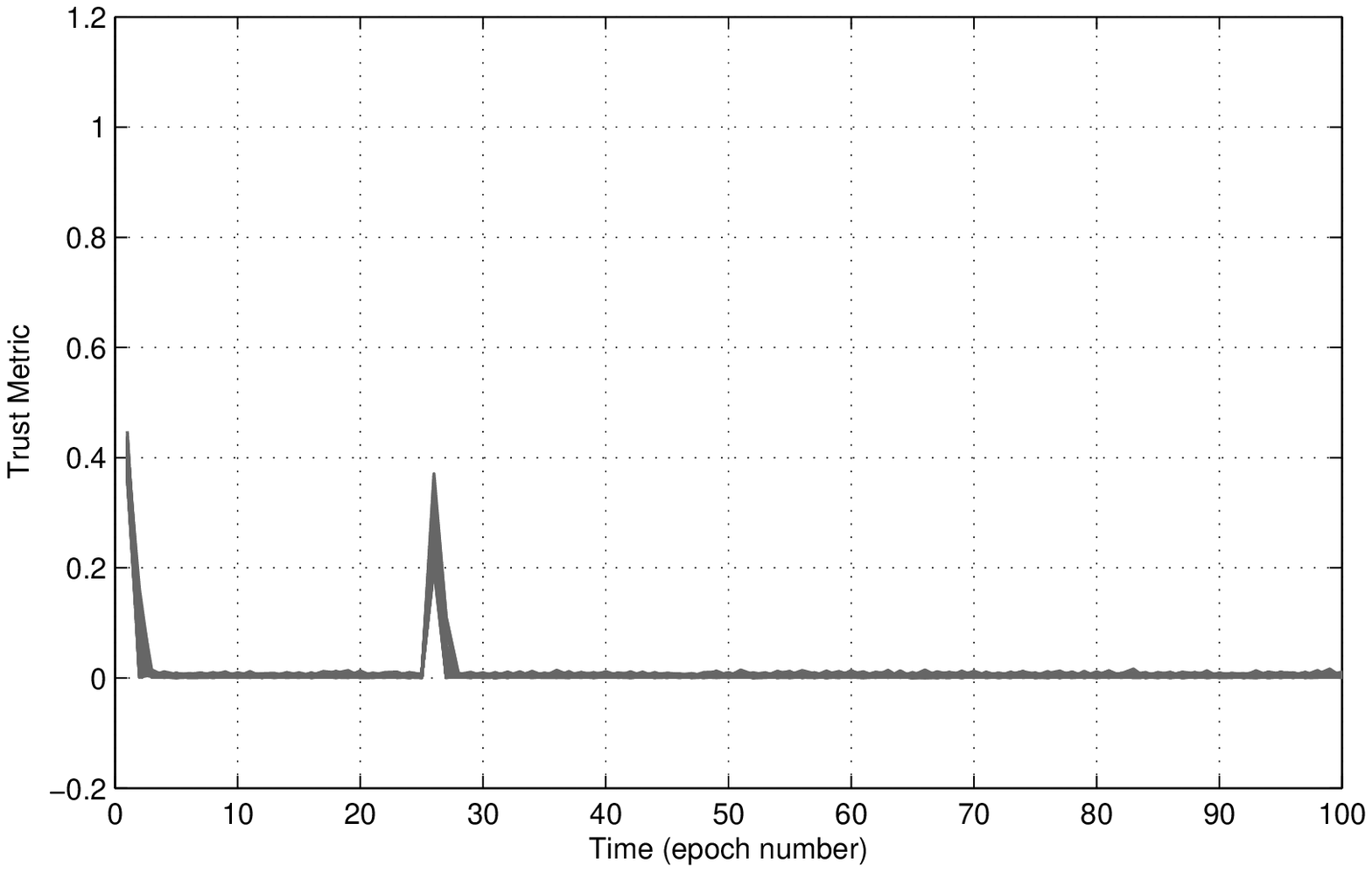}\includegraphics[width=3in,height=2.2in]{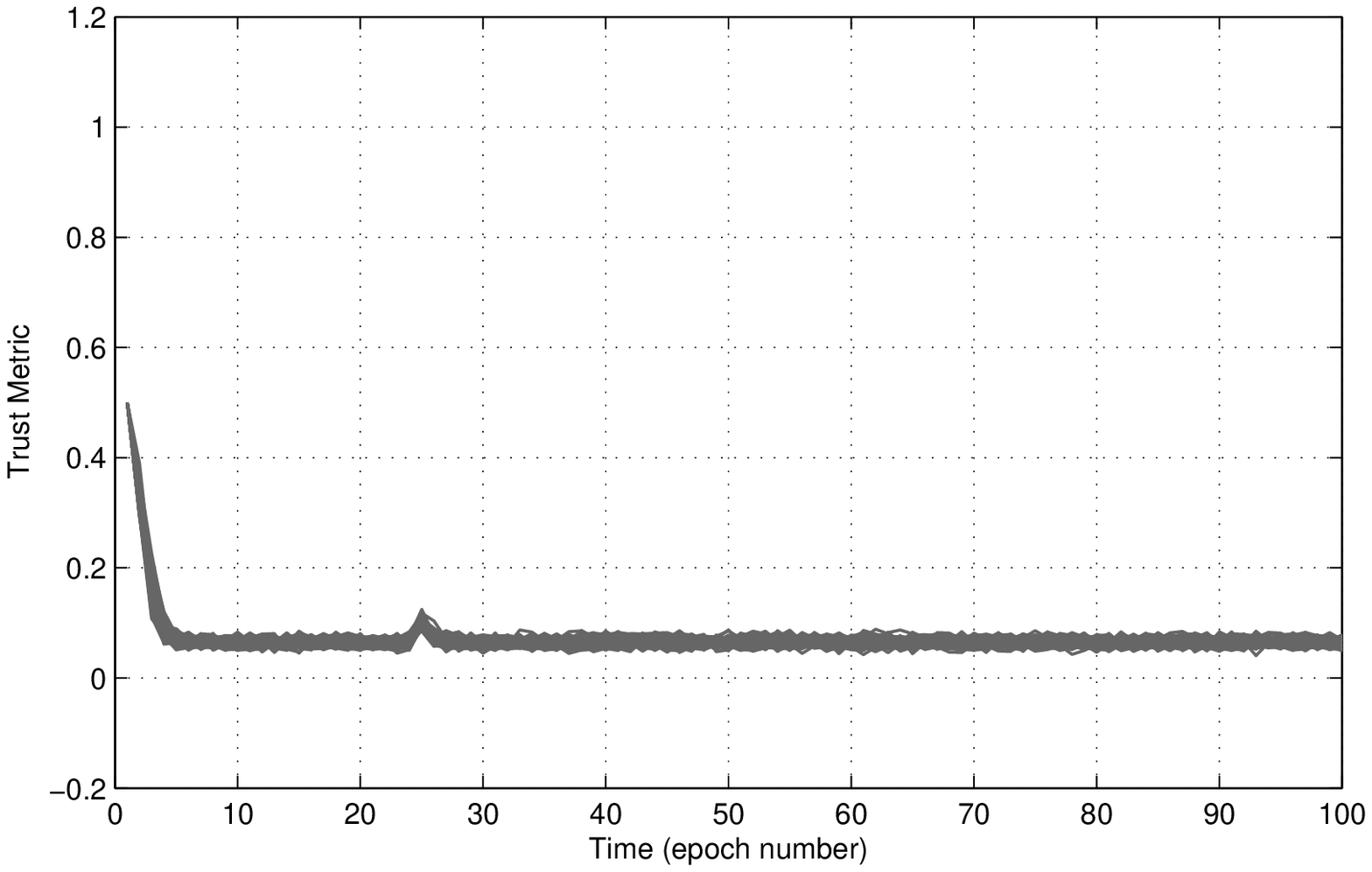}}
\end{tabular}
\caption{Left: Traces of the estimated trust metric of ``Sensor B" in 100 independent
Monte Carlo runs of the IPF algorithm. Right: Traces of the estimated trust metric of ``Sensor B" in 100 independent
Monte Carlo runs of the BDMPF algorithm.} \label{fig:Node_B_trust}
\end{figure}

\begin{figure}[H]
\begin{tabular}{c}
\centerline{\includegraphics[width=3in,height=2.2in]{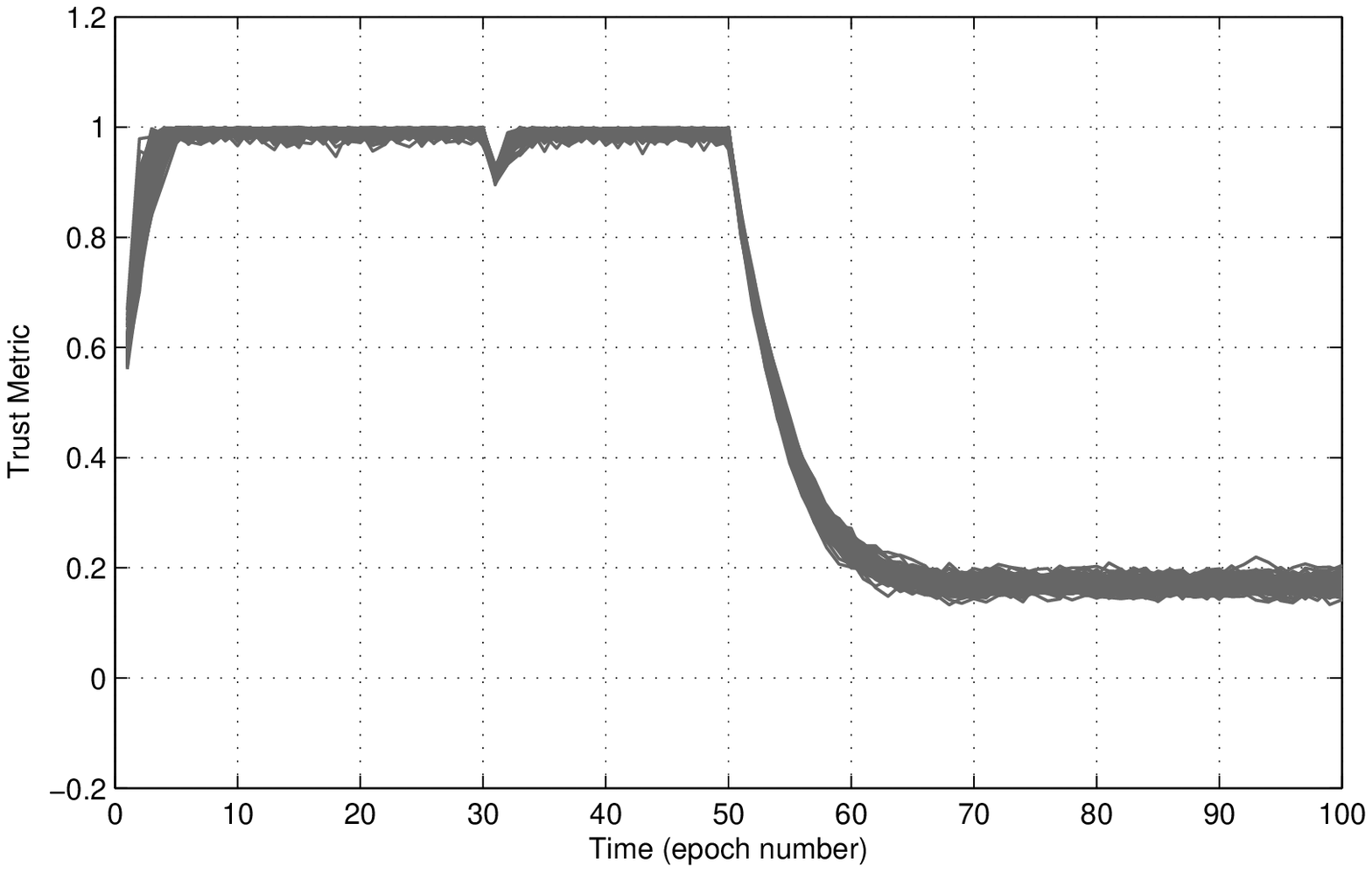}\includegraphics[width=3in,height=2.2in]{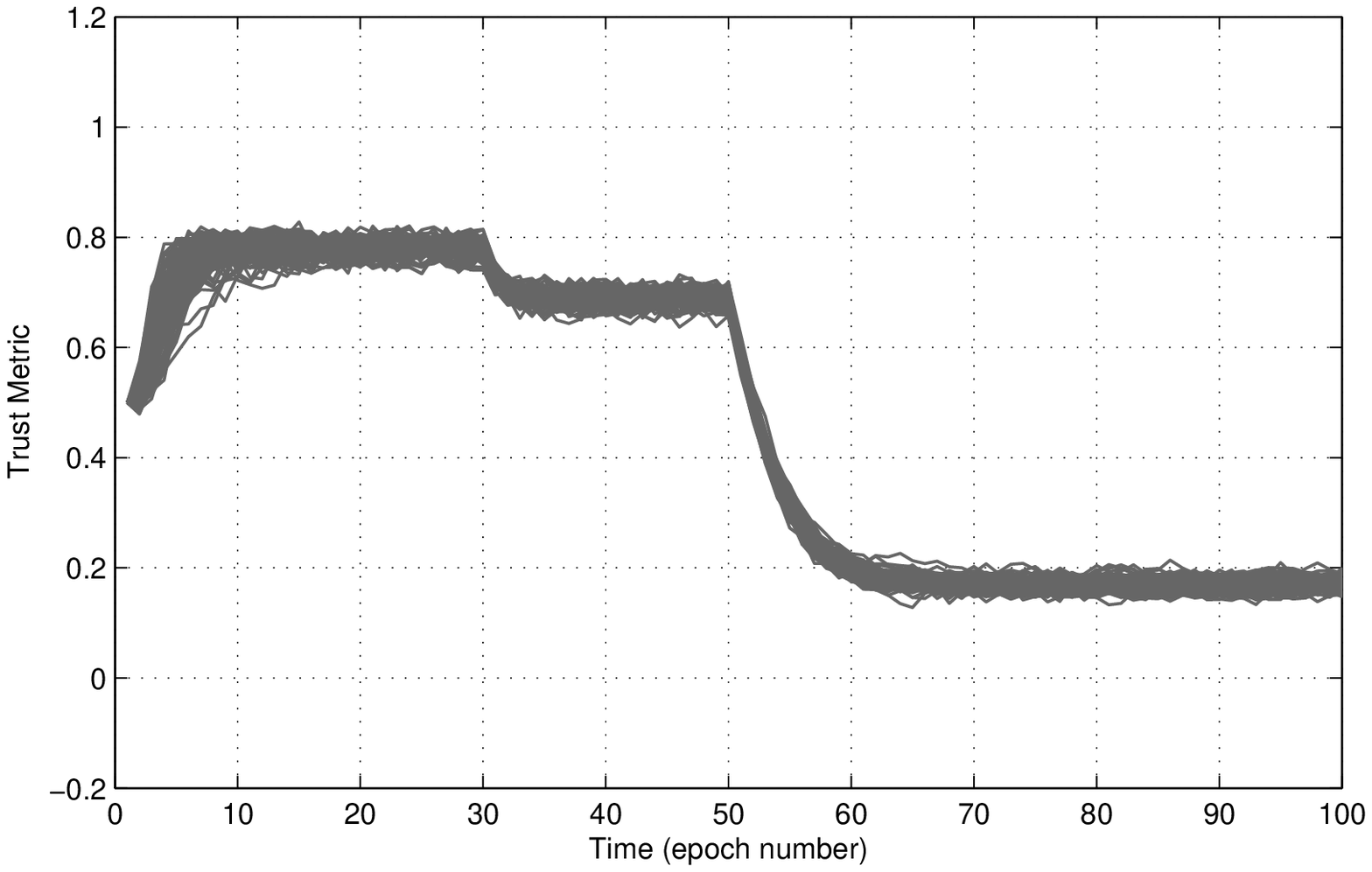}}
\end{tabular}
\caption{Left: Traces of the estimated trust metric of ``Sensor C" in 100 independent
Monte Carlo runs of the IPF algorithm. Right: Traces of the estimated trust metric of ``Sensor C" in 100 independent
Monte Carlo runs of the BDMPF algorithm.} \label{fig:Node_C_trust}
\end{figure}

\begin{figure}[!htb]
\begin{tabular}{c}
\centerline{\includegraphics[width=3in,height=2.2in]{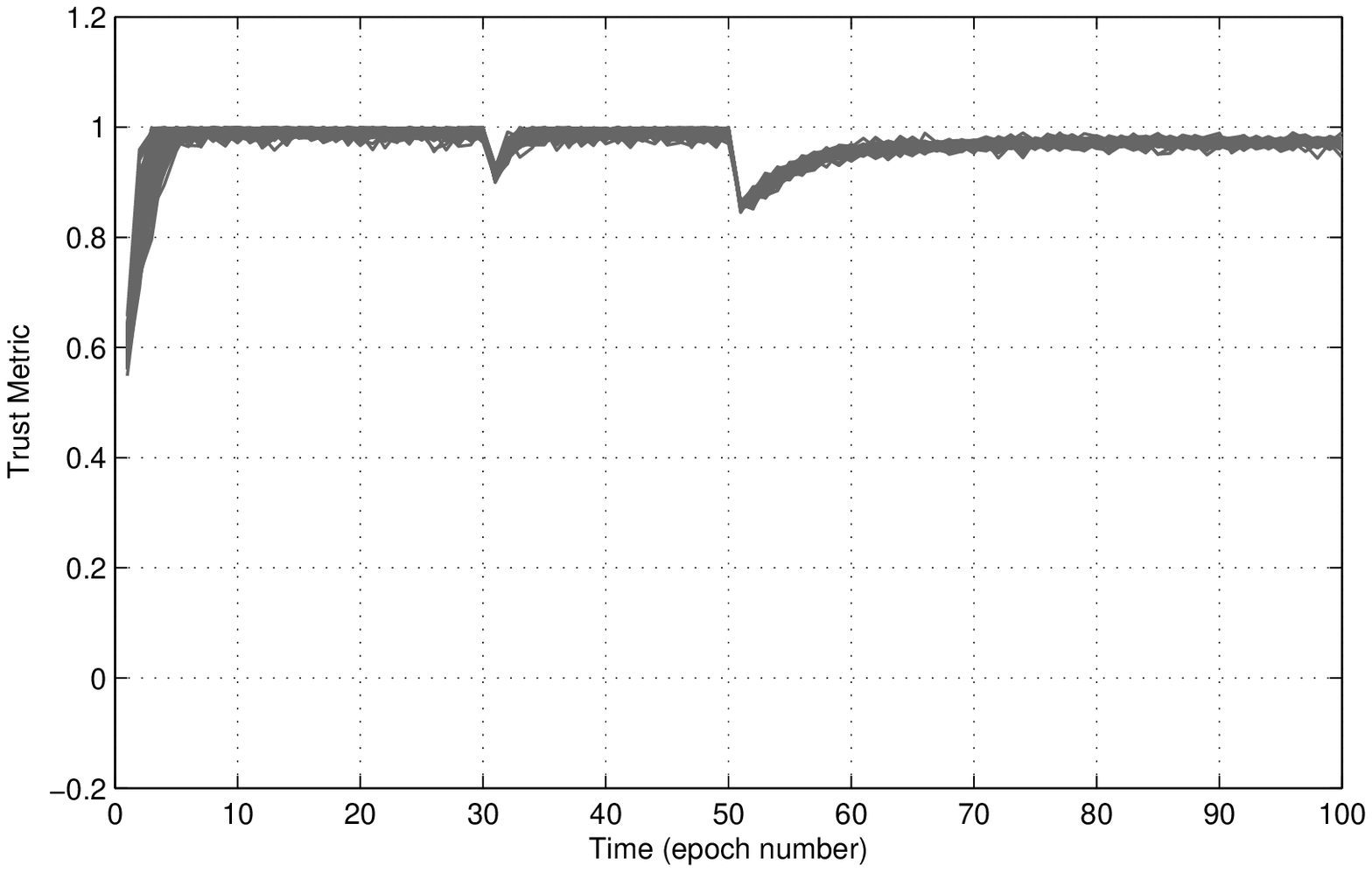}\includegraphics[width=3in,height=2.2in]{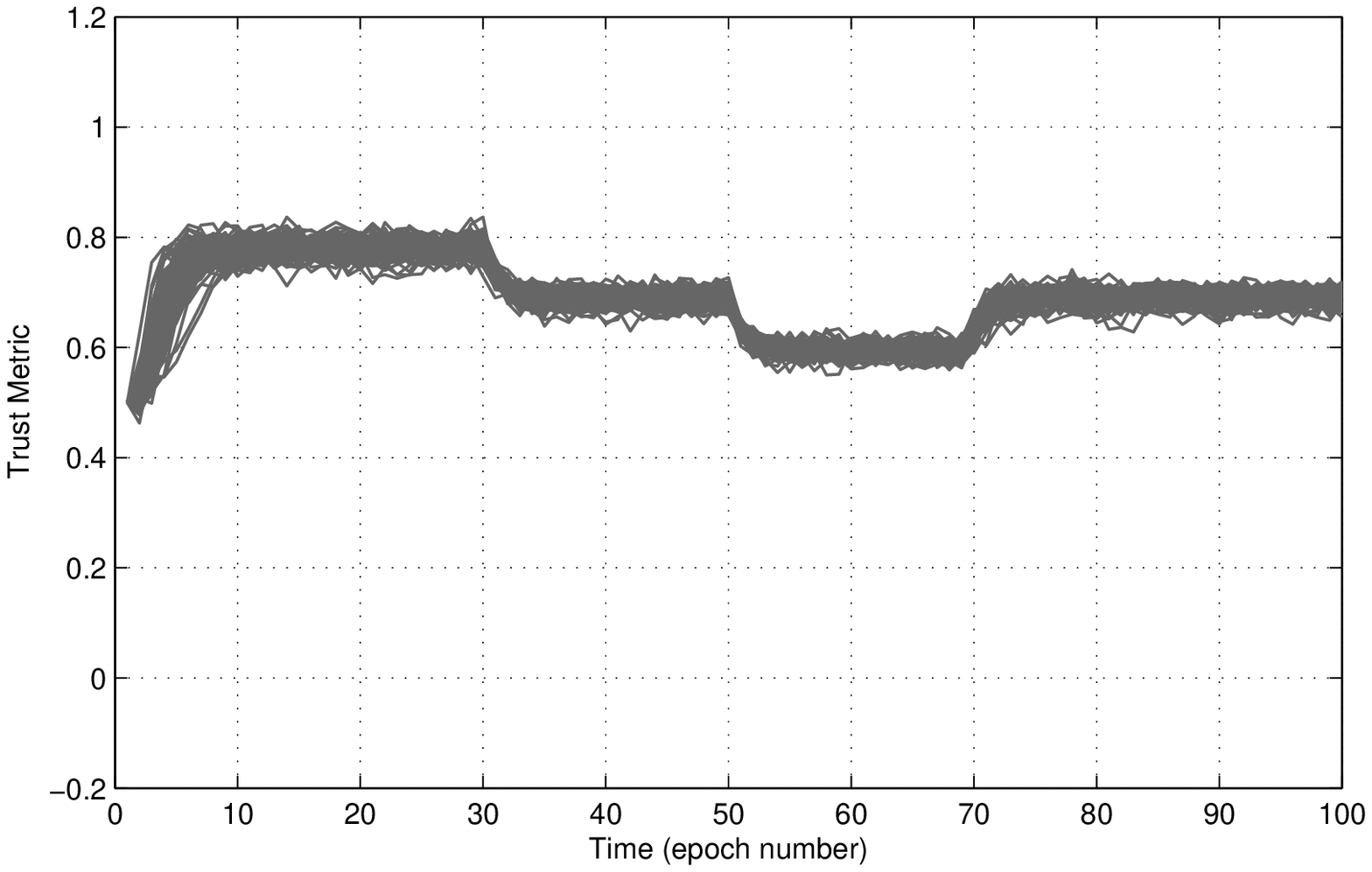}}
\end{tabular}
\caption{Left: Traces of the estimated trust metric of ``Sensor D" in 100 independent
Monte Carlo runs of the IPF algorithm. Right: Traces of the estimated trust metric of ``Sensor D" in 100 independent
Monte Carlo runs of the BDMPF algorithm.} \label{fig:Node_D_trust}
\end{figure}

The estimated traces of the trust metric of ``Sensor A", ``Sensor B", ``Sensor C" and
an always-trustworthy sensor node, termed ``Sensor D" here,
are plotted in Figs.\ref{fig:Node_A_trust}-\ref{fig:Node_D_trust}, respectively.

First let us analyze the estimation result on ``Sensor A".
According to the simulation setting described in Subsection \ref{sec:simulation_setting},
the trust metric of ``Sensor A" takes the value of 1
in two time periods, corresponding to the 1-30 and 71-100 time steps, and it takes the value of 0 at the other,
namely the 31-70 time steps.
In Fig.\ref{fig:Node_A_trust}, we see that within the first 10 time steps, the estimated trust metric of ``Sensor A"
given by the IPF algorithm converges to the
expected value 1 quickly, while the estimated trust metric given by the BDMPF algorithm converges to 0.8.
During the 31-70 time steps, the estimate given by the IPF algorithm converges again to the expected value 0 quickly,
while the BDMPF algorithm converges to a value close to 0.1. Regarding the last 30 time steps, it is shown that the IPF algorithm
can still output much accurate estimate on the trust metric,
while the performance of BDMPF deteriorates much more, since the gap between its estimate and the true answer is broadened.
The similar result of that the estimate given by IPF is much more accurate than that given by BDMPF can also be found in
Figs.\ref{fig:Node_B_trust}-\ref{fig:Node_D_trust}, for ``Sensors B, C and D", respectively.
Further, we can see that the performance gap between the BDMPF and the IPF algorithms is broadened when more sensor nodes
become untrustworthy.
For example, in Fig.\ref{fig:Node_D_trust}, we see that, for this always-trustworthy node ``Sensor D",
the estimated trust metric given by BMDPF
worsens along with the increase in the number of existing untrustworthy nodes. Specifically, we can see that at 51-70 time steps,
during which ``Sensors A, B and C" are all
untrustworthy, the BDMPF algorithm gives the worst estimate of the trust metric compared with in the other periods.
In contrast with BDMPF, the proposed IPF algorithm always provides an accurate estimate on the trust metric of ``Sensor D"
in Fig.\ref{fig:Node_D_trust}.
In another word, the IPF algorithm is shown to be remarkably much more robust than BDMPF in case of untrustworthy nodes being involved.
The above result is consistent with the theoretical analysis on the connections between the IPF and the BDMPF algorithms as
described in Subsection \ref{sec:connection}.
\subsubsection{Numerical performance evaluation}
For ease of quantitative performance evaluation, we used the root mean square error (RMSE) to measure the gap between the estimate provided by
an algorithm and the true answer. The RMSE regarding node $j$ at time step $k$ is defined to be
\begin{equation}
\mbox{RMSE}_{k,j}\triangleq\sqrt{\frac{\sum_{m=1}^M(\hat{x}_{k,j}^m-x_{k,j})^2}{M}},
\end{equation}
where $M$ denotes the total number of independent runs of the algorithm of our concern in the Monte Carlo simulation test,
$\hat{x}_{k,j}^m$ denotes the estimate of $x_{k,j}$ yielded in the $m$th independent run of the algorithm.
In what follows we set $M=100$.

\begin{figure}[!htb]
\begin{tabular}{c}
\centerline{\includegraphics[width=3in,height=1.8in]{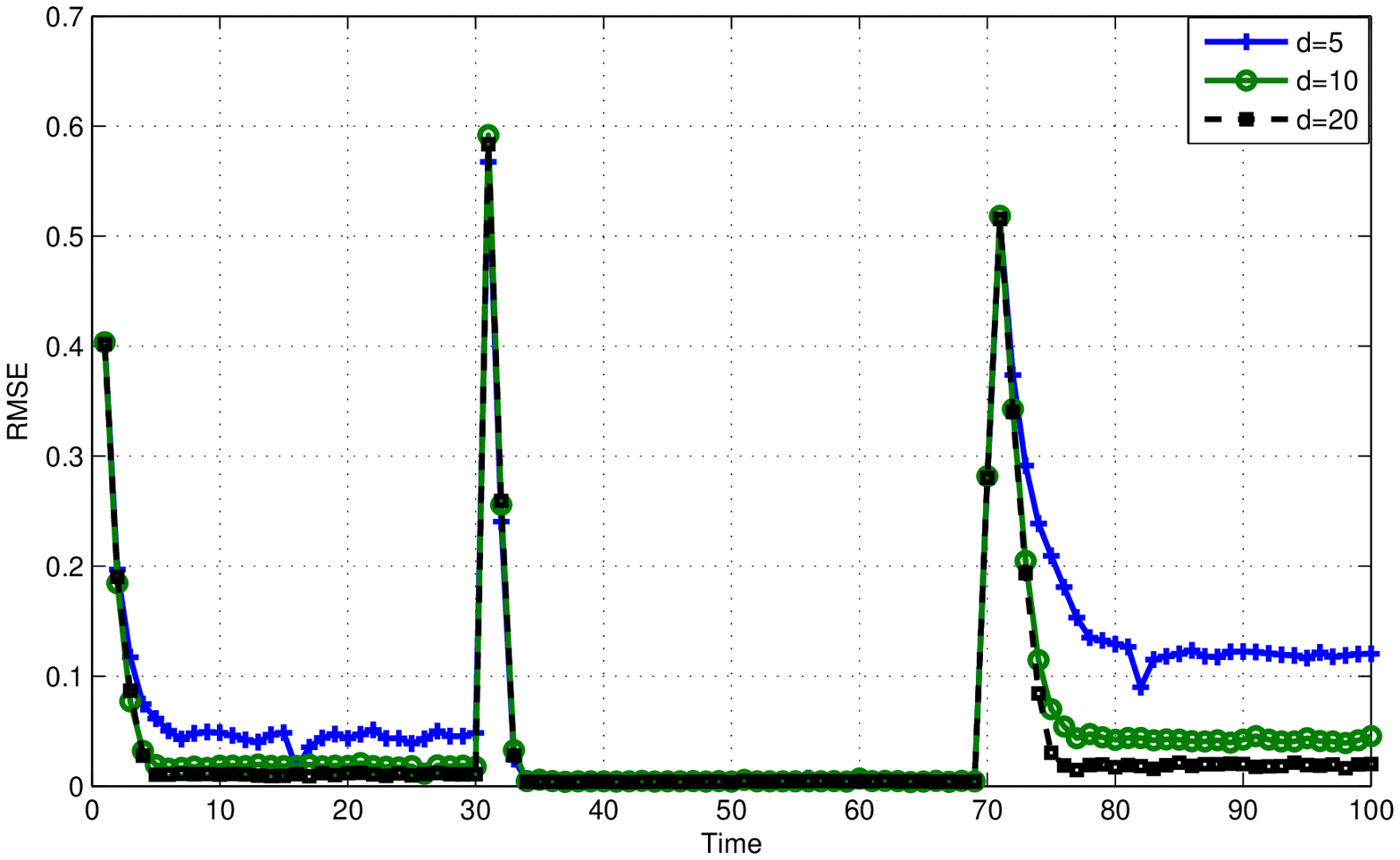}\includegraphics[width=3in,height=1.8in]{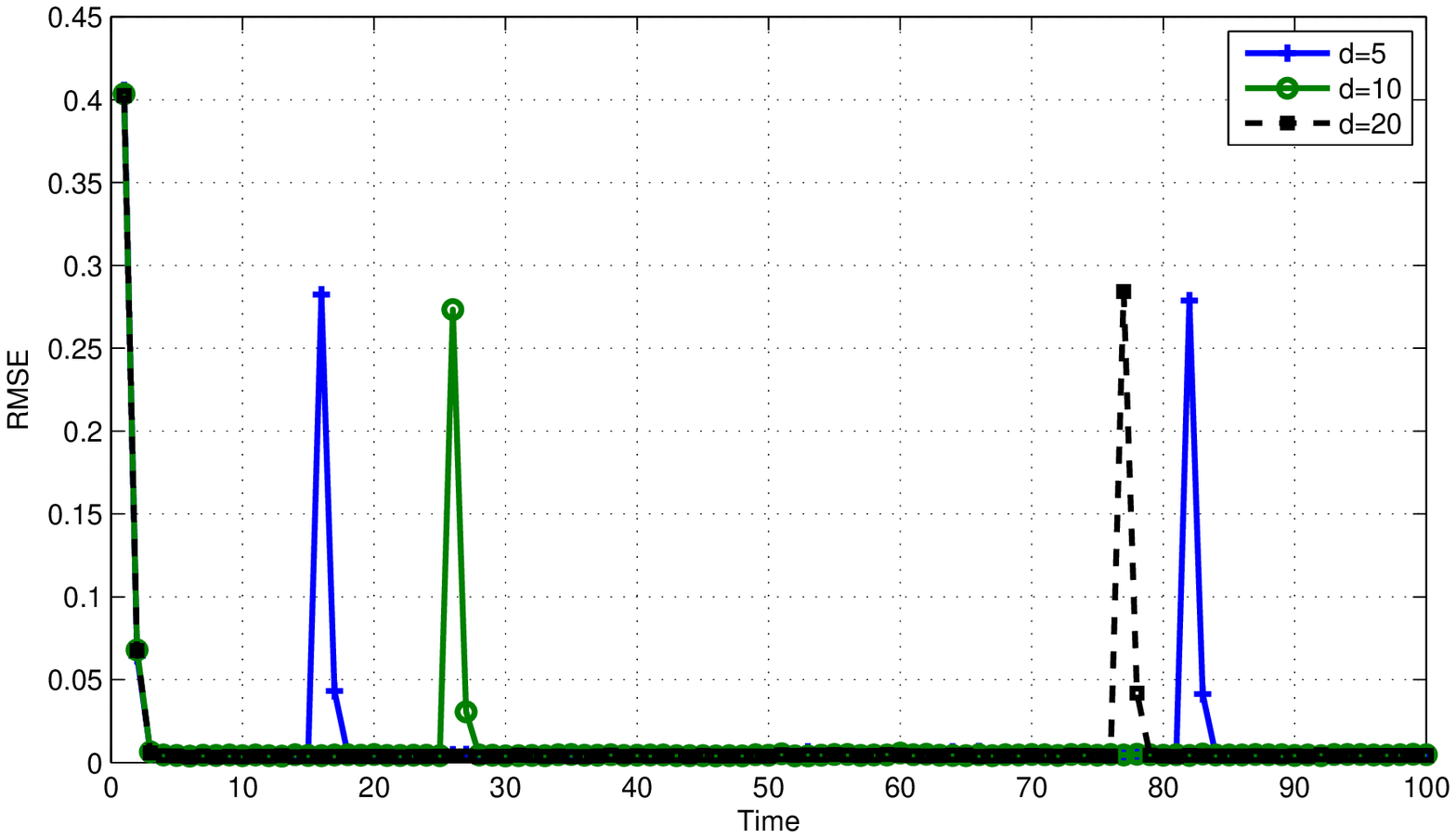}}\\
\centerline{\includegraphics[width=3in,height=1.8in]{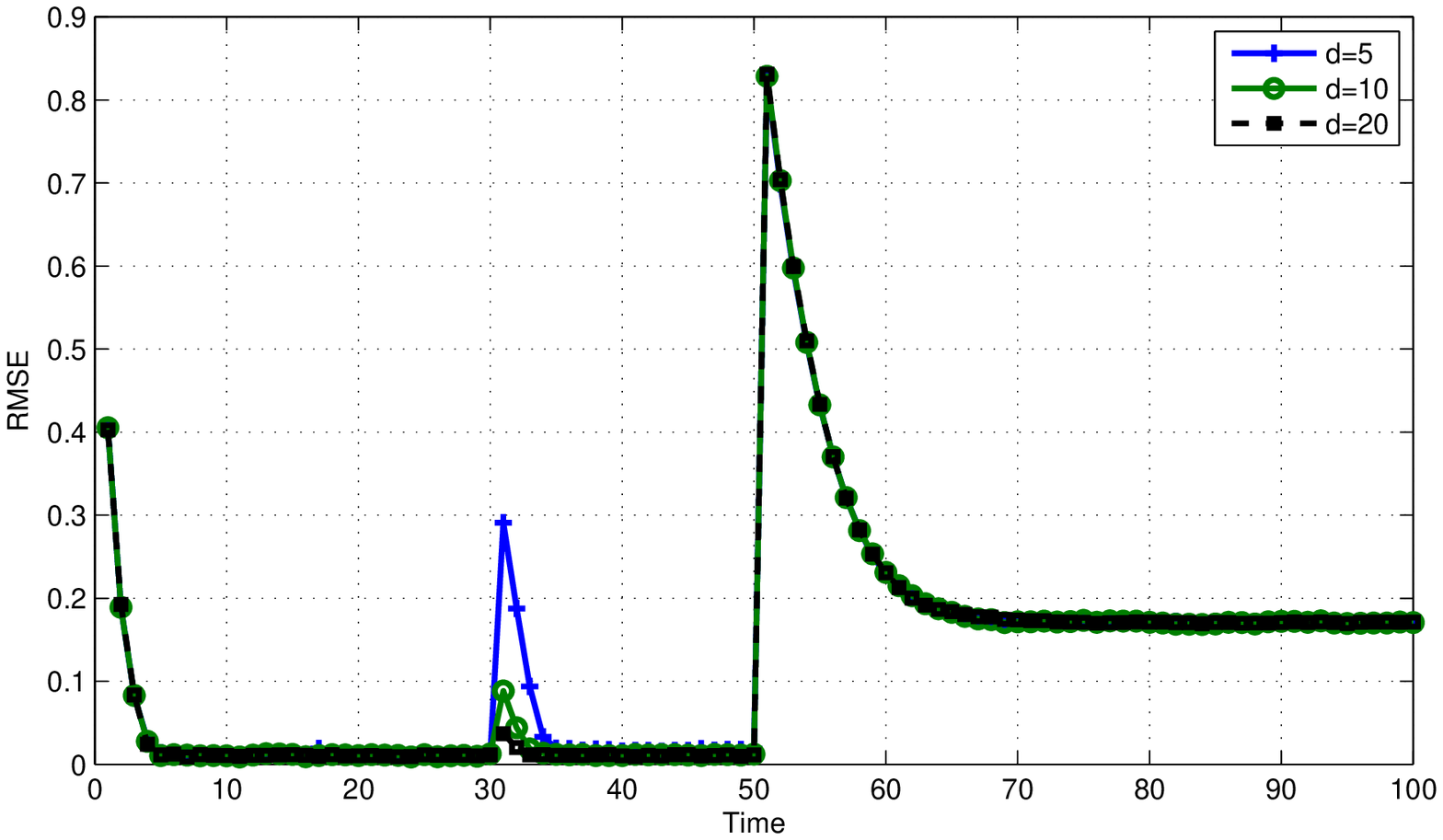}\includegraphics[width=3in,height=1.8in]{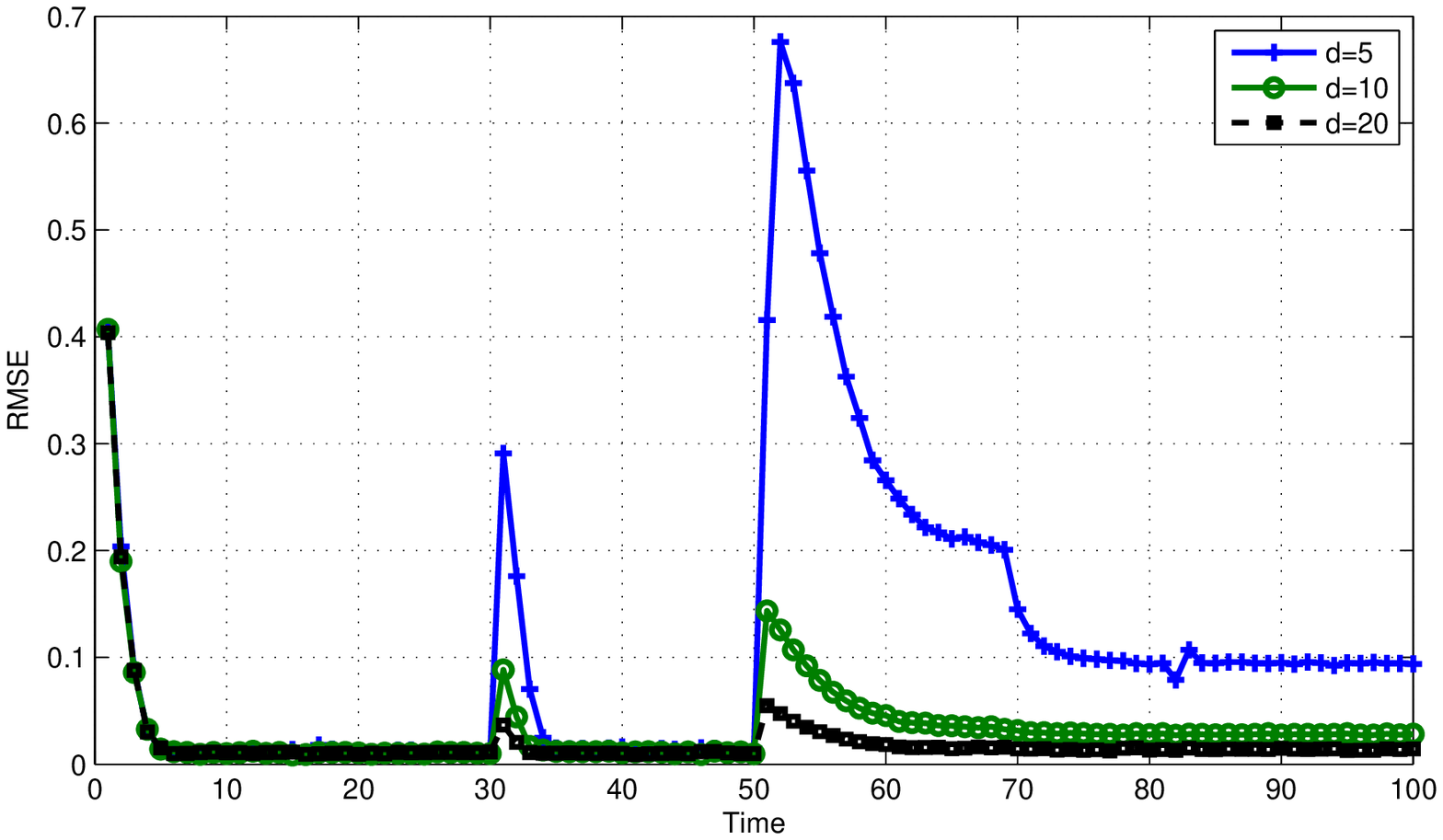}}
\end{tabular}
\caption{RMSE calculated based on simulation results obtained from 100 independent Monte Carlo runs of the IPF algorithm.
$d$ denotes the total number of sensor nodes under consideration.
The top left, top right, bottom left and bottom right sub-figures correspond to ``Sensors A, B, C and D", respectively.} \label{fig:rmse_simu}
\end{figure}

We investigated how the performance of IPF changes along with the dimension of the state $d$. We considered three cases,
corresponding to $d=5$, $d=10$ and $d=20$, respectively.
``Sensors A, B, C" with the same setting as before are involved for all cases.
For each specific $d$ value, we ran 100 independent Monte Carlo runs of the IPF algorithm, and then
calculated the corresponding RMSE.
The result is shown in Fig.\ref{fig:rmse_simu}, where we use ``Sensor D" to denote a representative always-trustworthy node
as before.
It is shown that, as $d$ gets bigger, the RMSE gets smaller, and, even in case of $d=5$, most of the time the RMSE does not exceed $0.12$.

\begin{figure}[H]
\begin{tabular}{c}
\centerline{\includegraphics[width=3in,height=1.8in]{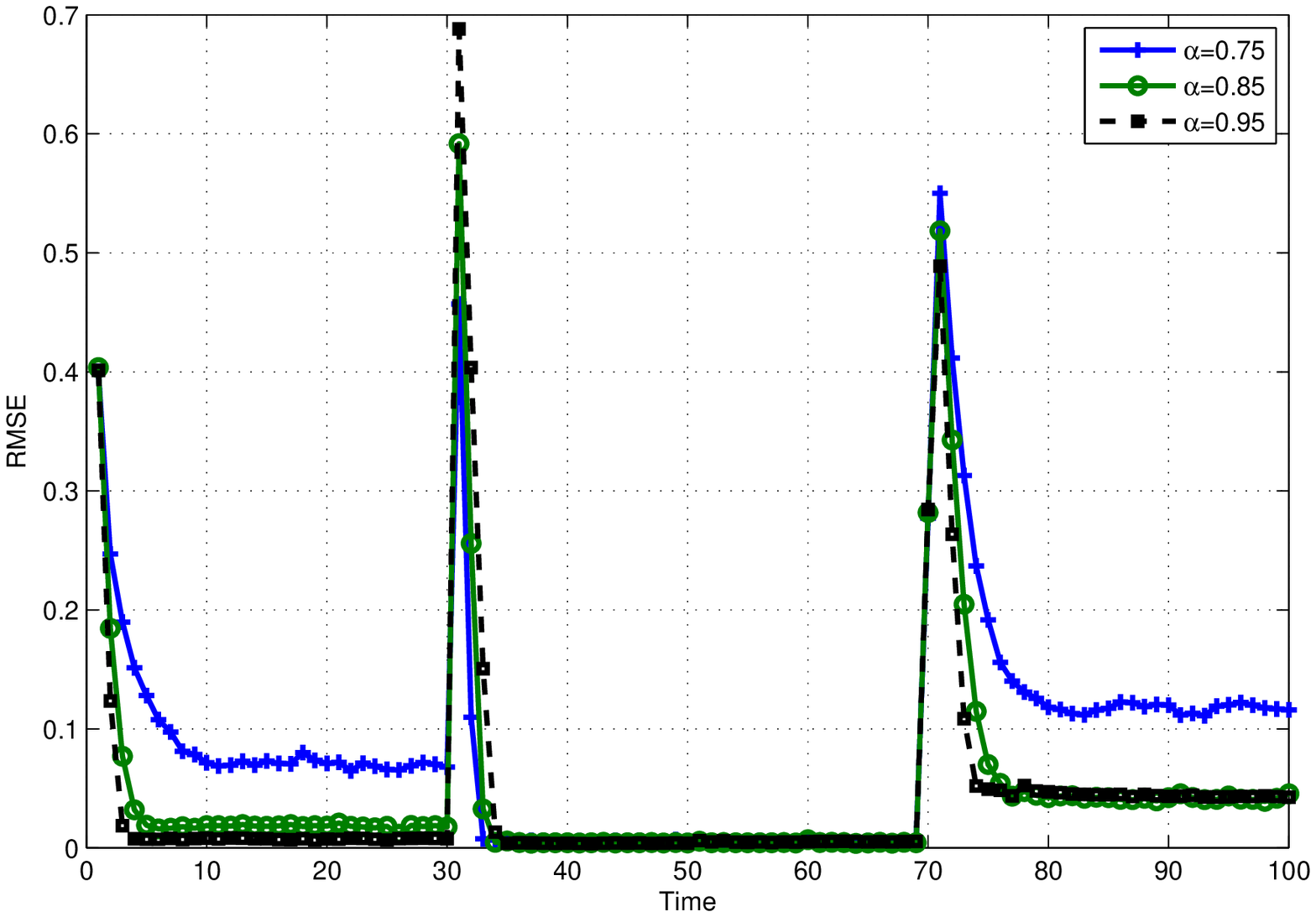}\includegraphics[width=3in,height=1.8in]{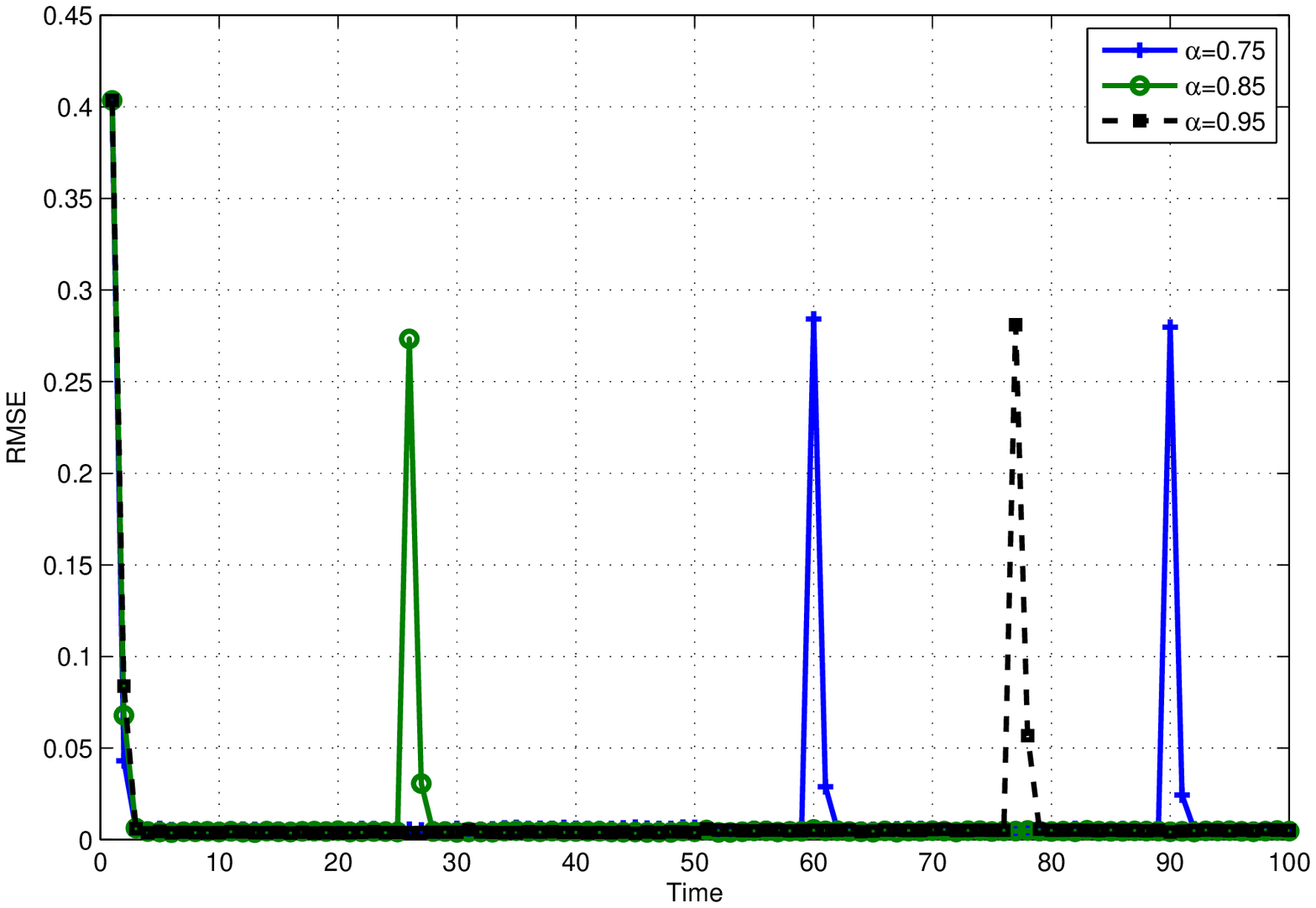}}\\
\centerline{\includegraphics[width=3in,height=1.8in]{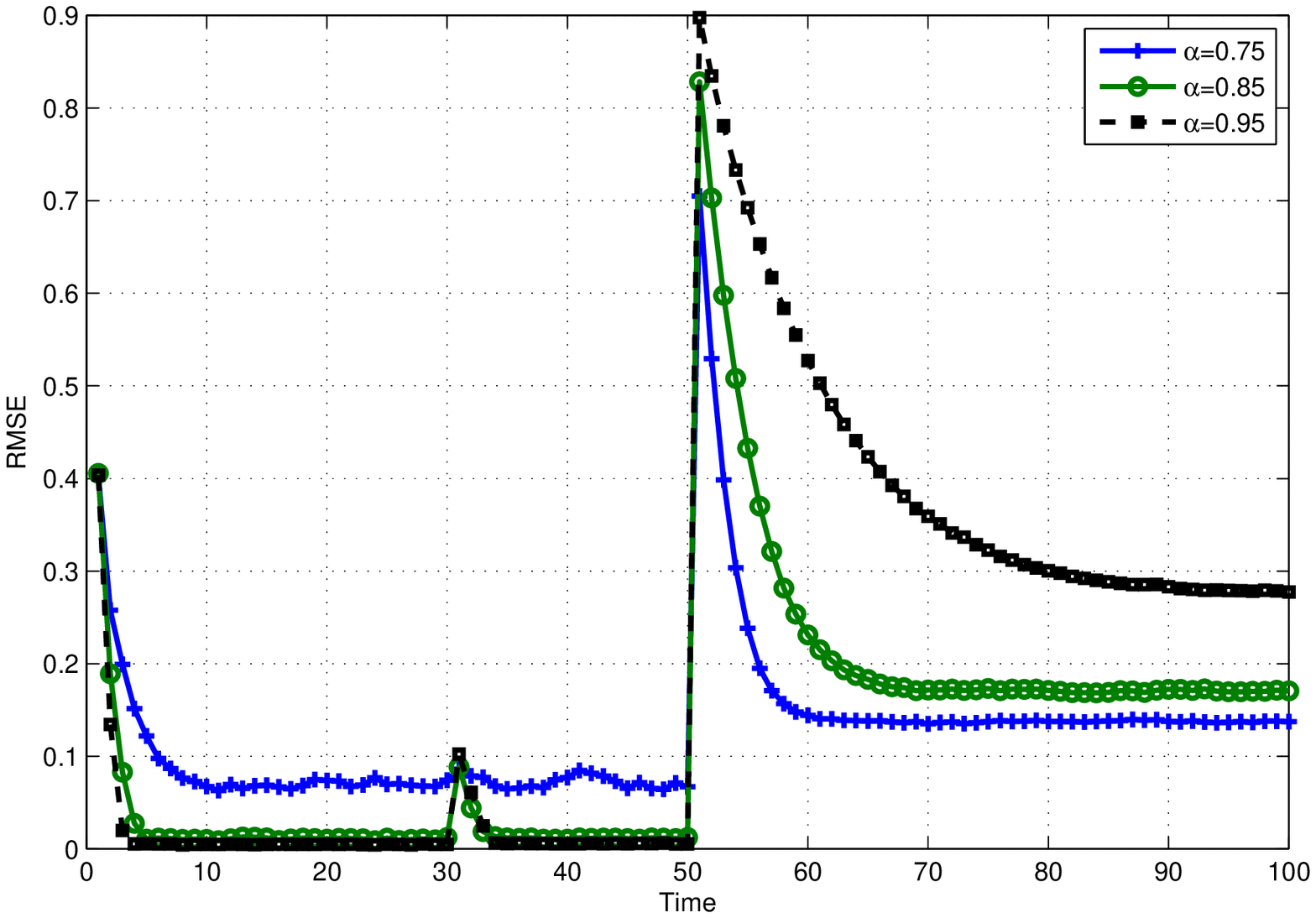}\includegraphics[width=3in,height=1.8in]{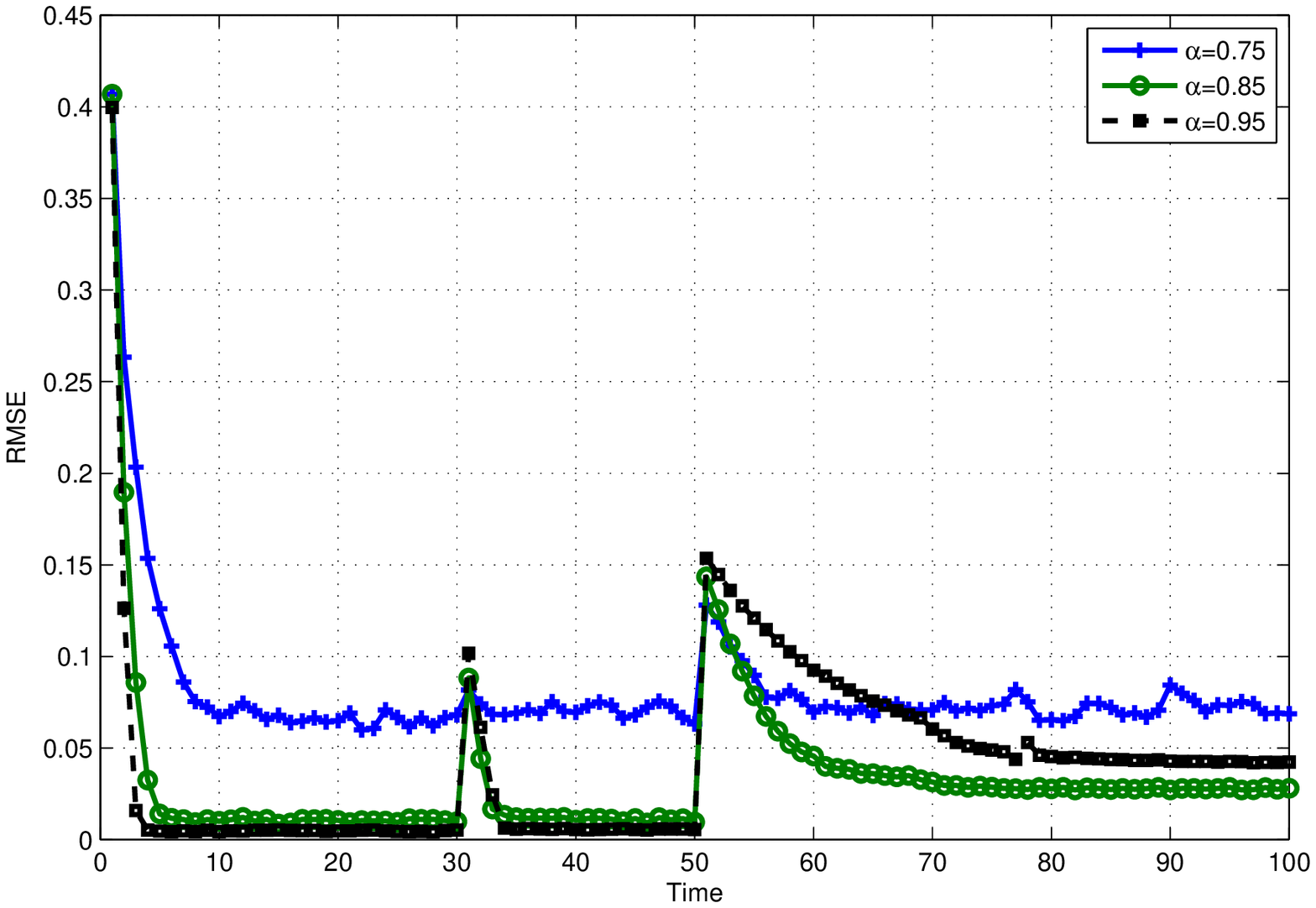}}
\end{tabular}
\caption{RMSE calculated based on simulation results obtained from 100 independent Monte Carlo runs of the IPF algorithm under cases
with different $\alpha$ values.
The top left, top right, bottom left and bottom right sub-figures correspond to ``Sensors A, B, C and D", respectively.} \label{fig:rmse_alpha}
\end{figure}

We also investigated the influence of the aging parameter $\alpha$ in Eqn.(\ref{eqn:state_trans}) on the performance of the IPF algorithm.
We considered three cases corresponding to $\alpha=0.75$, $\alpha=0.85$ and $\alpha=0.95$, respectively. For each case, we set $d=10$, and ran
100 independent Monte Carlo runs of the IPF algorithm. The results are shown in Fig.\ref{fig:rmse_alpha}.
We see that, most of the time, the RMSE corresponding to $\alpha=0.85$ and $\alpha=0.95$ is smaller than that corresponding to $\alpha=0.75$, for all the
sensor nodes under consideration. In the bottom left sub-figure, we see that 0.85 is more preferable to 0.75 in initializing $\alpha$.
Actually 0.85 is selected empirically as the default value of $\alpha$ in our algorithm.

\subsubsection{Investigation on the iterative component and the computational burden of the IPF algorithm}
\begin{figure}[H]
\begin{tabular}{c}
\centerline{\includegraphics[width=3in,height=1.8in]{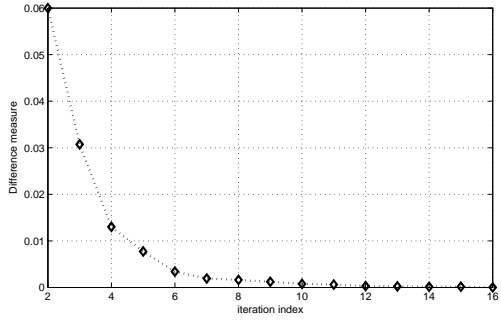}}\\
\end{tabular}
\caption{Convergence of the component-wise inference procedure in the IPF algorithm. The X and Y label of the figure denote the iteration index $m$ and $\sqrt{\|\hat{x}_k-X_o\|/d}$ in Algorithm \ref{algo:iteration_IPF}, respectively.} \label{fig:convergence_IPF}
\end{figure}

The proposed IPF algorithm includes an iterative process, namely the the component-wise inference procedure, as shown in Algorithm \ref{algo:iteration_IPF}, while our experiments show that it needs just a few iterations in order to converge, see Fig.\ref{fig:convergence_IPF}.
The computational time of the IPF algorithm in three cases, corresponding to $d=5$, $d=10$ and $d=20$, respectively, is presented in Table 2.
So experimentally, we see that the computational burden of the IPF algorithm is linearly related with the dimension of the state $d$. Such a good scaling property of our algorithm is especially desirable when dealing with high dimensional cases.
\begin{table}[H]
\caption{Elapsed time of an independent run of the IPF algorithm. T$_{elapsed}$ denotes
 the real value of the elapsed time. T$_{scaled}$ denotes the scaled version of T$_{elapsed}$, calculated on the basis of the
 5 dimensional case.}
\begin {center}
\begin{tabular}{c|c|c|c}
\hline\hline
$d$ & 5 & 10 & 20\\
\hline
T$_{elapsed}$ (unit:second) & 71.5 & 124.6 & 281.3\\
\hline
T$_{scaled}$ & 1 & 1.7 & 3.9\\
\hline\hline
\end{tabular}
\end {center}
\end{table}
\subsection{Real data analysis results}
\begin{figure}[H]
\begin{tabular}{c}
\centerline{\includegraphics[width=3.5in,height=2.5in]{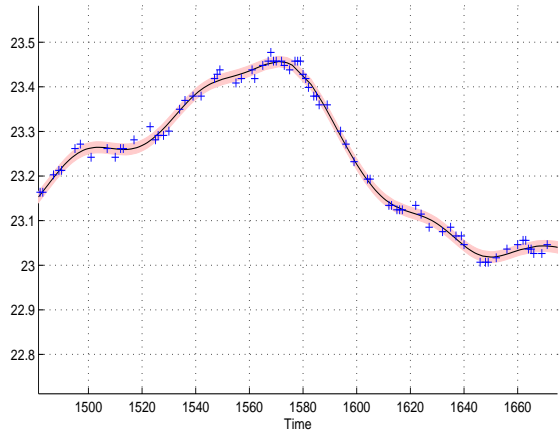}}\end{tabular}
\caption{Fitting sensor readings with Gaussian Process regression. The shadow depicts the one standard error uncertainty associated with the fitted curve based on the sensor readings, represented by the plus signs.} \label{fig:GP_fit}
\end{figure}

Here we describe an evaluation performed on the Intel Lab Data \cite{intellabdata}, a public data set collected from 54 sensors
deployed in the Intel Berkeley Research lab between February 28th and April 5th, 2004. In our experiment, we chose the whole day's data from February 28th, remaining only the sensor reading attribute of
original data set, i.e. humidity, temperature, and light.
We selected a spatial neighbor set of sensors 9, 10, 11, 12, 13 for analysis. As the sampling time of the sensor readings
reported by different sensors is not synchronous, we performed Gaussian process regression \cite{rasmussen2010gaussian} to
fit the readings of each sensor. A snapshot of the fitting effect is depicted in Fig.\ref{fig:GP_fit}.

We adopted the fault models described in \cite{ganeriwal2008reputation} to simulate faulty sensor readings, which are then
injected into the original data. The purpose is to evaluate whether the IPF algorithm can detect such faults in time
through estimating the trust metric of each sensor online.
Specifically, for the 1st node under consideration, we removed its reported data between the 500th and 700th epoch to
simulate the phenomenon termed ``Sleeper Attacks" \cite{ganeriwal2008reputation}.
For the 2nd node, we modified its sensor readings
between the 300th and 400th epoch to be a constant 100. This phenomenon is called ``Stuck-at Fault" in \cite{ganeriwal2008reputation}.
For the 3rd node, we added a zero-mean Gaussian noise with standard error 20, to each of its sensor readings
between the 200th and 250th epoch, and this is the so-called ``Variance Degradation Fault" described
in \cite{ganeriwal2008reputation}. For the 4th node, we added an offset value, 100, to its
pre-fault measurement values between the 100th to the 150th epoch with a probability 0.5. This type of fault is termed ``offset fault"
in \cite{ganeriwal2008reputation}.
The IPF algorithm is initialized by the same parameter values as shown in Table 1, except that we empirically set $r=2$ here.
\begin{figure}[H]
\begin{tabular}{c}
\centerline{\includegraphics[width=3in,height=1.8in]{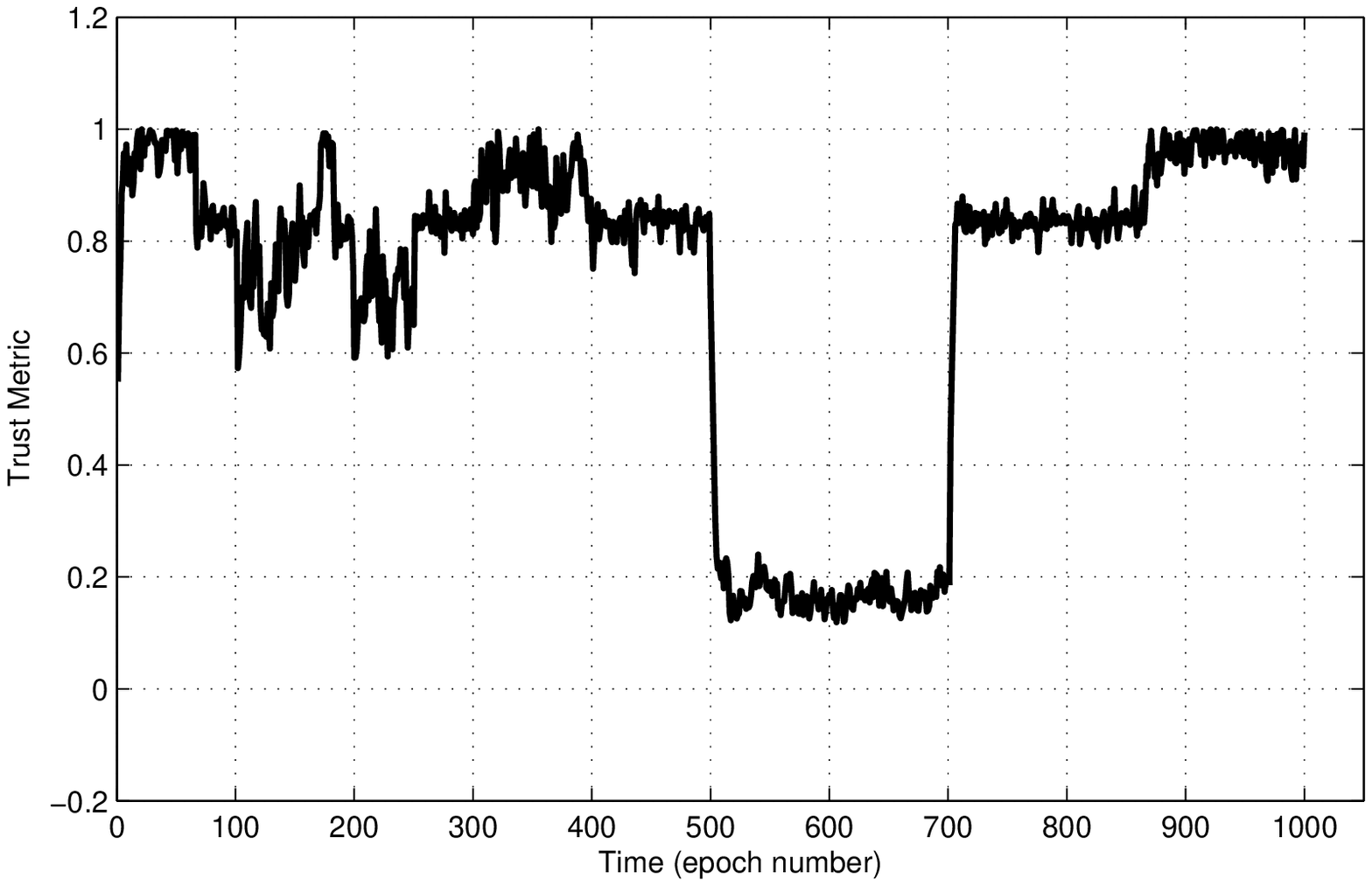}\includegraphics[width=3in,height=1.8in]{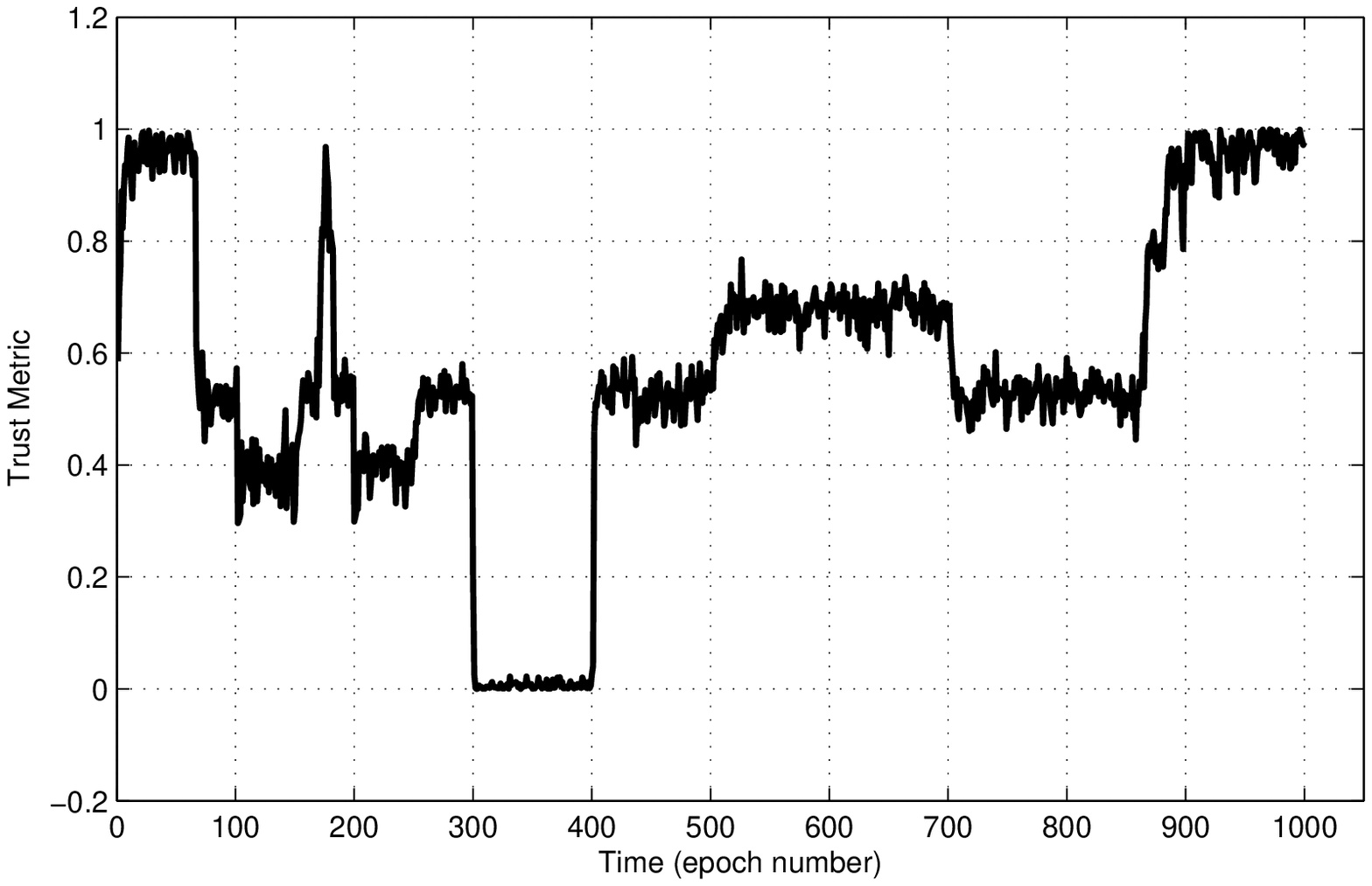}}\\
\centerline{\includegraphics[width=3in,height=1.8in]{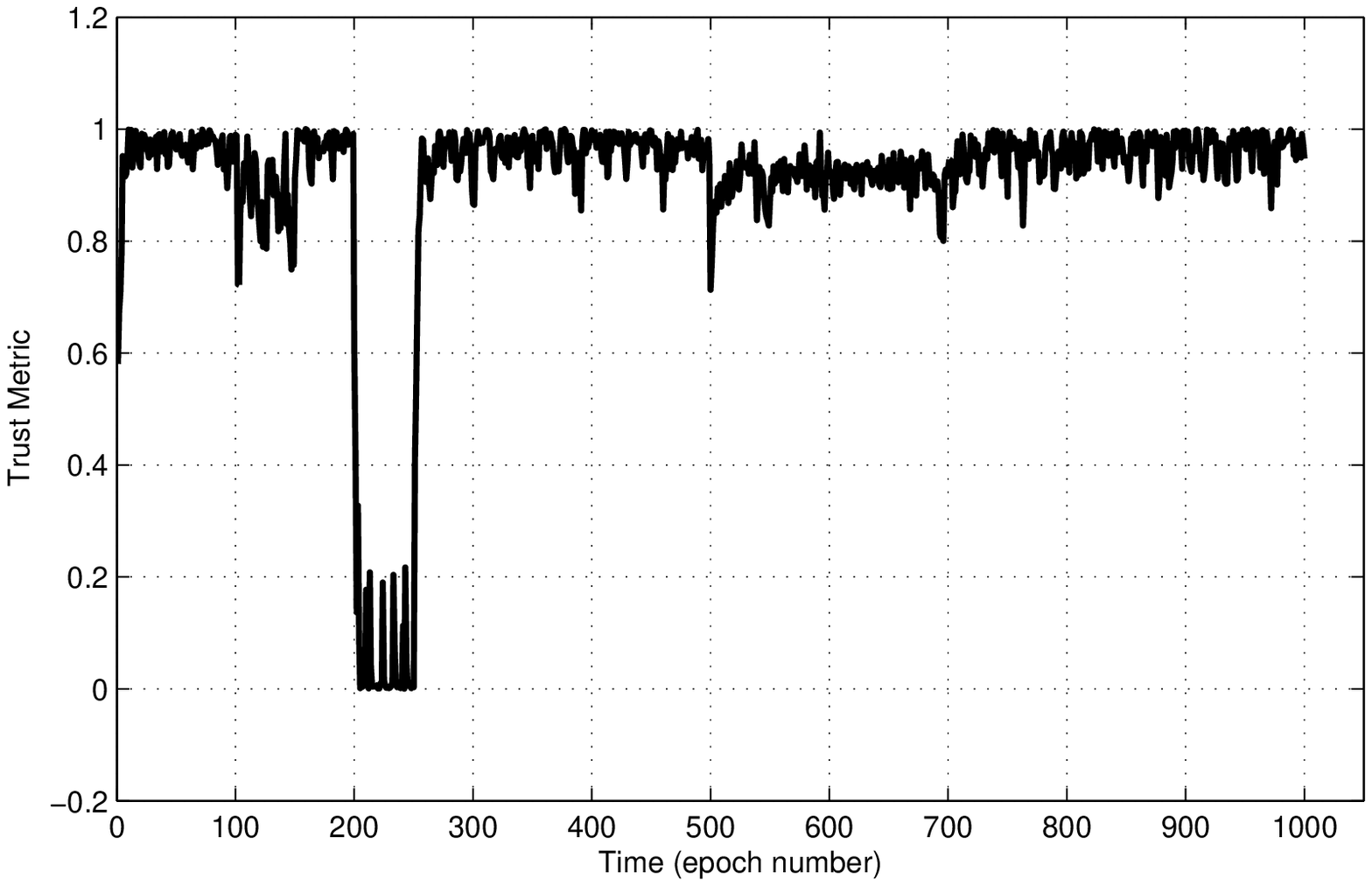}\includegraphics[width=3in,height=1.8in]{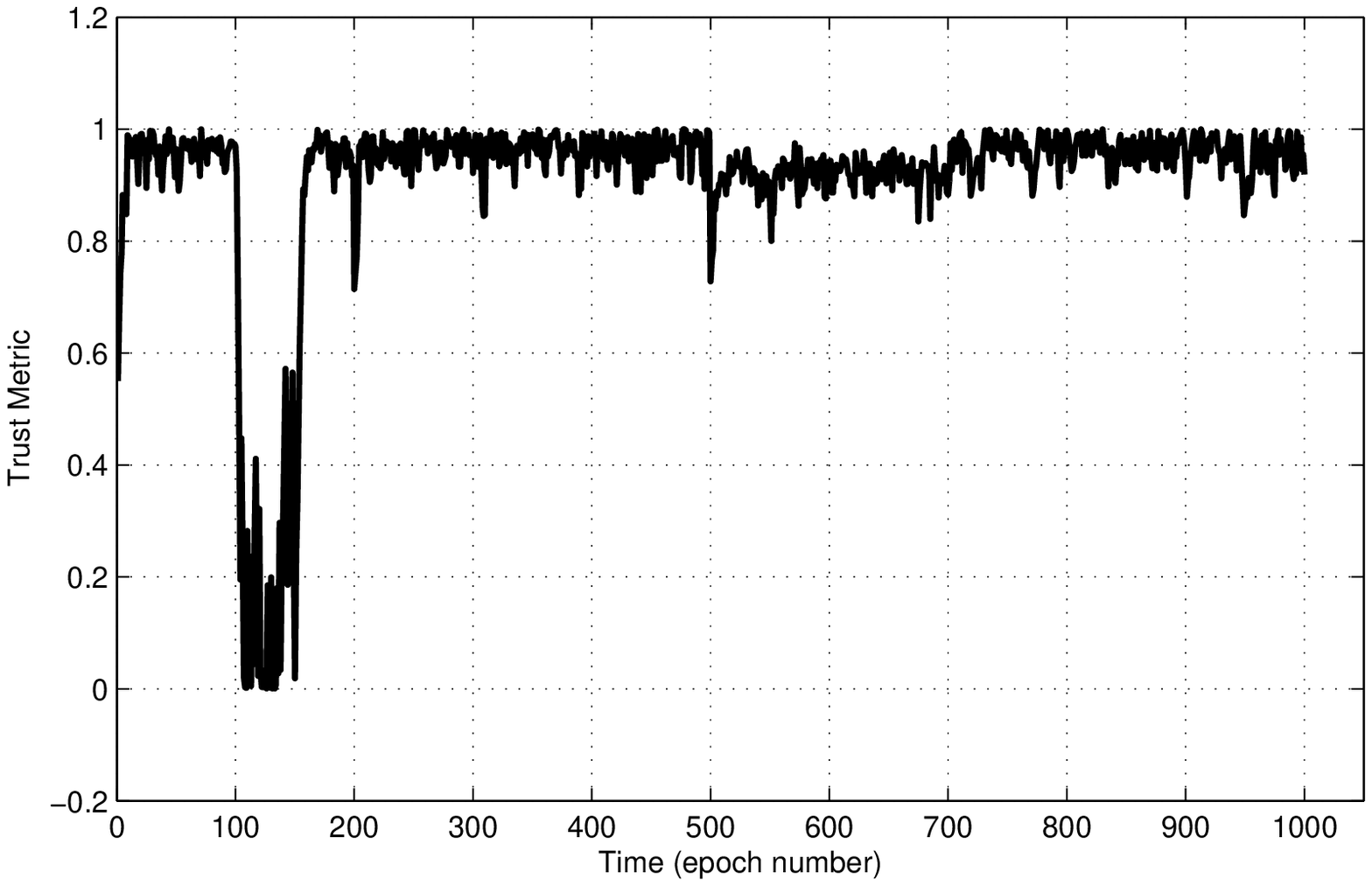}}
\end{tabular}
\caption{Trust evaluation in presence of faults in the Intel lab data. The top left, top right, bottom left and bottom right sub-figures correspond to
the 1st sensor node with ``Sleeper Attacks" between the 500th and 700th epoch, the 2nd sensor node with ``Stuck-at Fault" between the 300th and
400th epoch, the 3rd sensor node with ``Variance Degradation Fault" between the 200th and
250th epoch, and the 4th sensor node with ``offset fault" between the 100th and
150th epoch, respectively.} \label{fig:trust_intel_data}
\end{figure}
The estimated trust metric for the above-mentioned sensor nodes is graphically presented in Fig.\ref{fig:trust_intel_data}. As is shown,
the estimated trust metric, given by our IPF algorithm, can accurately reflect the existence of different types of faults online.
Thus the IPF algorithm can be regarded as an efficient fault detection tool.
\section{Conclusions}
In this paper, we propose a theoretical data-driven modeling framework to address the problem of trust evaluation over WSN.
The basic idea is to treat the problem of trust evaluation from the perspective of nonlinear state filtering.
In particular, we design a generic trust model, termed SSTM. Then, making use of the information on the model structure, we design a corresponding state filtering algorithm, termed IPF. Through both extensive simulation studies and real data analysis, we evaluated the performance of the IPF algorithm. The results show that it can yield accurate estimate on the trust metric of the sensor nodes online,
even in complex environments, wherein different types of non-trustworthy nodes exist and report different types of faulty measurements to the server node.
The computational complexity of the proposed algorithm is shown to be linearly related with the dimension of the state. Such a scaling property makes our algorithm easy to meet the practical constraints in energy, memory, and computation power, especially when we have a lot of sensor nodes waiting to be evaluated concurrently. The future work is planned to compare the proposed algorithm with alternatives in the aspects of consumptions in energy, memory, and computation power. In addition, by virtue of Bayesian decision making theory, the proposed framework here can be generalized to deal with risk analysis and decision making issues.
\section{Acknowledgments}
This work was partly supported by the National Natural Science Foundation
of China (NSFC) under grant No. 61571238, China Postdoctoral Science Foundation under grant Nos. 2015M580455 and 2016T90483, China Postdoctoral International Academic Exchange Program and National NSFC under grant No. 91646116.

\ifCLASSOPTIONcompsoc
\ifCLASSOPTIONcaptionsoff
  \newpage
\fi
\bibliographystyle{IEEEtran}
\bibliography{mybibfile}
\end{document}